\documentclass[preprint,10pt]{aastex}
\usepackage{grffile}

\usepackage{url}                                                                                  

\newcommand{\anchorfoot}[2] {\anchor{#1}{#2}\footnote{\url{#1}}}
\newcommand{\anchorparen}[2]{\anchor{#1}{#2} (\url{#1})}

\usepackage{hyperref}
\renewcommand{\anchor}[2]{\href{#1}{#2}}

\newcounter{PSUfootnotemark}
\newcounter{HRMAfootnotemark}

\newcommand{\Chandra}{{\em Chandra}}
\newcommand{\ACIS}   {{ACIS}}
\newcommand{\CIAO}   {{\em CIAO}}
\newcommand{\Sherpa} {{\em Sherpa}}
\newcommand{\Chart}  {{\em ChaRT}}

\newcommand{\DSnine} {{\em DS9}}
\newcommand{\MARX}   {{\em MARX}}
\newcommand{\AEacro} {{\em AE}}
\newcommand{\TARA}   {{\em TARA}}
\newcommand{\XSPEC}  {{\em XSPEC}}

\newcommand{\IDL}    {{\em IDL}}
\newcommand{\Racro}  {{\em R}}

\newcounter{column_number}
\newcommand{\numberthecolumn}{\colhead{(\arabic{column_number})}\stepcounter{column_number}}

\shorttitle{Innovations}
\shortauthors{Broos et al.}                                
\slugcomment{Accepted by the \apj, 2010 Mar 10 (\#343576)}

\begin{document}

\title{Innovations in the Analysis of \Chandra-\ACIS\ Observations}

\author{
Patrick S. Broos,\altaffilmark{1} 
Leisa K. Townsley,\altaffilmark{1} 
Eric D. Feigelson,\altaffilmark{1} 
Konstantin V. Getman,\altaffilmark{1} 
Franz E. Bauer,\altaffilmark{2}\altaffilmark{3}
Gordon P. Garmire\altaffilmark{1} 
}
\email{patb@astro.psu.edu}

\altaffiltext{1}{Department of Astronomy \& Astrophysics, 525 Davey Laboratory, 
Pennsylvania State University, University Park, PA 16802, USA} 

\altaffiltext{2}{Space Science Institute, 4750 Walnut Street, Suite 205, Boulder, Colorado 80301, USA}
\altaffiltext{3}{Pontificia Universidad Cat\'{o}lica de Chile, Departamento de Astronom\'{\i}a y Astrof\'{\i}sica, Casilla 306, Santiago 22, Chile}

\begin{abstract}

As members of the instrument team for the Advanced CCD Imaging Spectrometer (\ACIS) on NASA's {\it Chandra X-ray Observatory} and as \Chandra\ General Observers, we have developed a wide variety of data analysis methods that we believe are useful to the \Chandra\ community, and have constructed a significant body of publicly-available software (the  {\it ACIS Extract} package) addressing important \ACIS\ data and science analysis tasks.
This paper seeks to describe these data analysis methods for two purposes: to document the data analysis work performed in our own science projects, and to help other \ACIS\ observers judge whether these methods may be useful in their own projects (regardless of what tools and procedures they choose to implement those methods).

The \ACIS\ data analysis recommendations we offer here address much of the workflow in a typical \ACIS\ project, including data preparation, point source detection via both wavelet decomposition and image reconstruction, masking point sources, identification of diffuse structures, event extraction for both point and diffuse sources, merging extractions from multiple observations, nonparametric broad-band photometry,  analysis of low-count spectra, and automation of these tasks.
Many of the innovations presented here arise from several, often interwoven, complications that are found in many \Chandra\ projects: large numbers of point sources (hundreds to several thousand), faint point sources, misaligned multiple observations of an astronomical field, point source crowding, and scientifically relevant diffuse emission.

\end{abstract}


\section{INTRODUCTION \label{introduction.sec}}

Since its launch in 1999, the {\it Chandra X-ray Observatory} \citep{Weisskopf02} has revolutionized X-ray astronomy.
\Chandra\ provides remarkable angular resolution---unlikely to be matched by another X-ray observatory within the next two decades---and its most commonly used instrument, the Advanced CCD Imaging Spectrometer (\ACIS), produces observations with a very low background \citep{Garmire03}.\footnote{See also the \anchorparen{http://asc.harvard.edu/proposer/POG/pog_pdf.html}{Chandra Proposers' Observatory Guide}.}
These two technical capabilities allow detection of point sources with as few as ${\sim}5$ observed X-ray photons (commonly referred to as ``events'' or ``counts''), a data analysis regime unique among X-ray observatories.  
Observations of Galactic star clusters and mosaics of nearby galaxies or extragalactic deep fields often produce hundreds to thousands of weak X-ray sources. 
\Chandra's excellent sensitivity to point sources and angular resolution also provide a unique capability for studying diffuse emission superposed onto those point sources, since they can be effectively identified and then masked.

\begin{figure}[htb]
\centering
\includegraphics[width=1.0\textwidth]{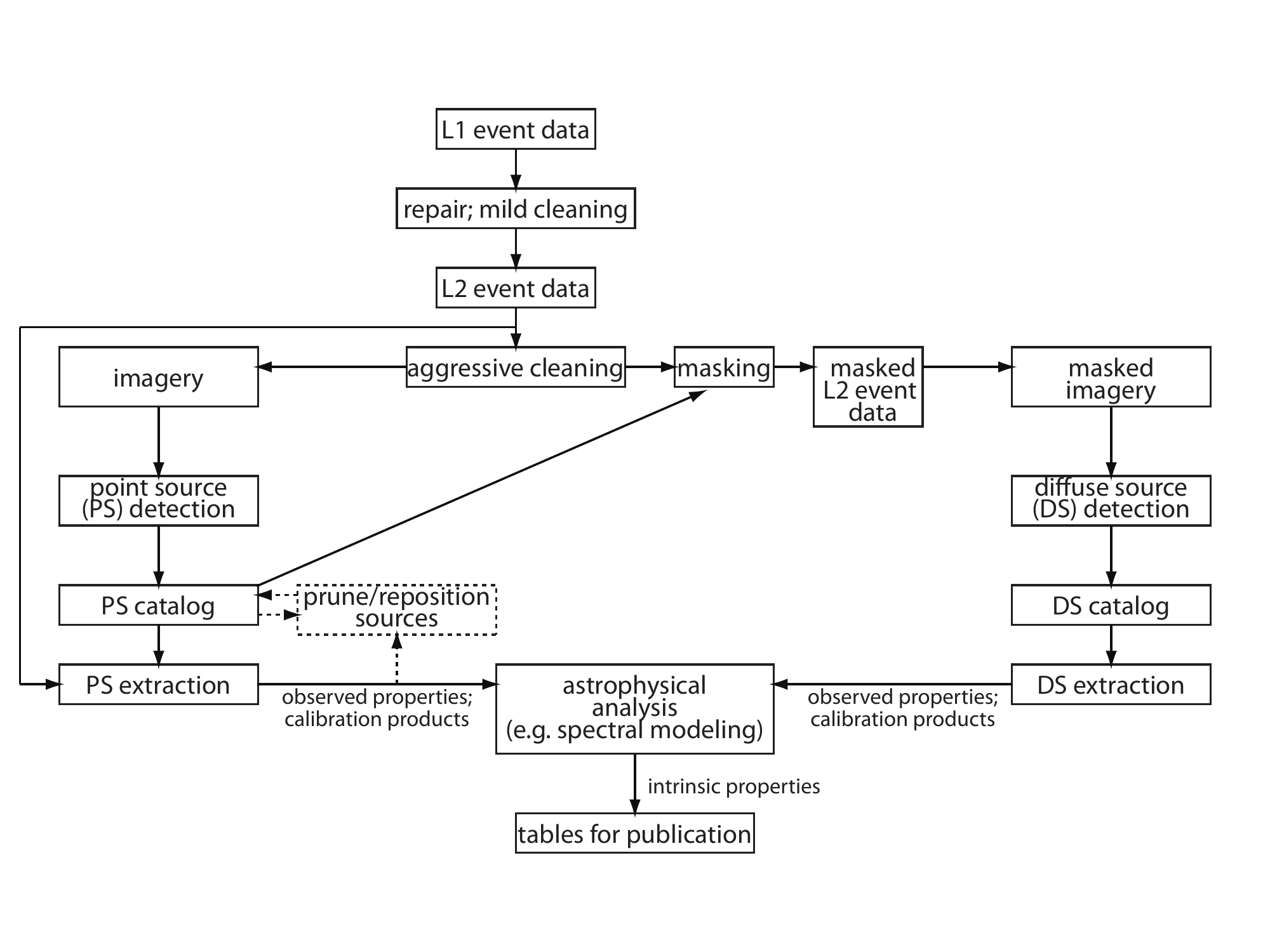}
\caption{The data analysis workflow described here for a single \ACIS\ field containing multiple point sources (PS, left branch) and diffuse sources (DS, right branch).  
\label{workflow.fig}}
\end{figure}

For many types of \ACIS\ ``imaging'' studies,\footnote{
Our discussion here is limited to data taken in the most common \ACIS\ configuration, called Timed Exposure Mode, at either the \ACIS-I or \ACIS-S aimpoint.  Dispersed data from the \Chandra\ gratings are not addressed here.
} 
most observers follow a data analysis workflow that is similar to that outlined in Figure~\ref{workflow.fig}.
Relatively raw data derived from satellite telemetry, known as 
\anchorfoot{http://cxc.harvard.edu/ciao/dictionary/levels.html}{``Level 1 Data Products''} (L1), are passed through a variety of repair and cleaning operations to produce 
``Level 2 Data Products'' (L2) that are appropriate for analysis.
A common workflow for studying point sources (solid boxes and arrows on left side of Figure~\ref{workflow.fig}) consists of binning the 
L2 event data into one or more images that are searched for sources.
The events and background associated with each point source in the catalog are ``extracted'' and calibrated.
Observed sources properties (e.g., count rates, apparent fluxes, spectra, light curves) are estimated, then combined with calibration products to estimate intrinsic astrophysical source properties.
A common and similar workflow for studying diffuse emission (right side of Figure~\ref{workflow.fig}) consists of removing (``masking'') the point sources from the data, constructing images, identifying several regions of diffuse emission to study, and then extracting and analyzing those diffuse sources in a manner similar to that used for point sources.

\begin{figure}[htb]
\centering
\includegraphics[width=0.8\textwidth]{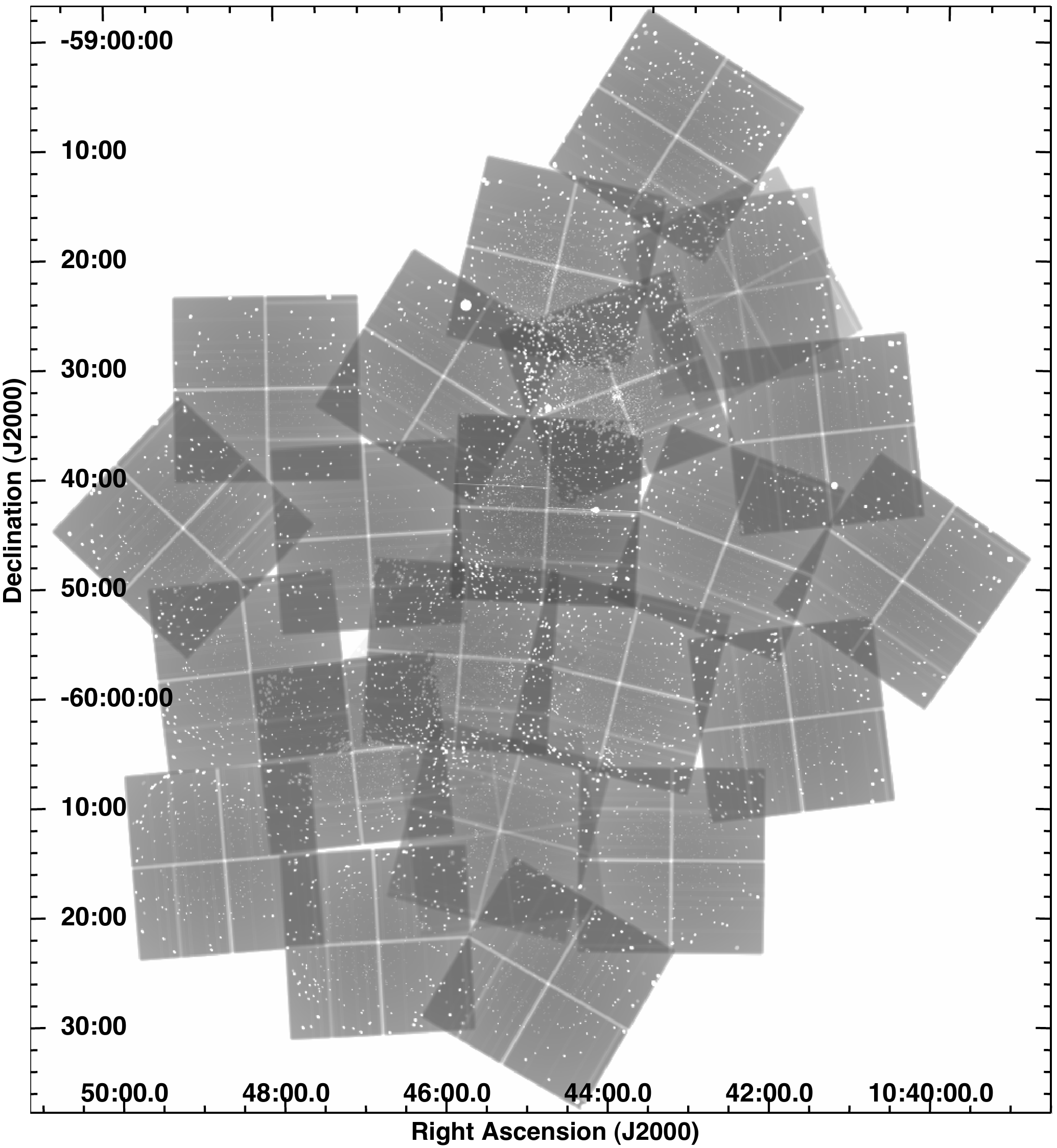}
\caption{
Exposure map for the Chandra Carina Complex Project \citep{Townsley10} study of the Carina Nebula comprised of 22 ACIS-I pointings (38 observations), with point sources masked (\S\ref{bkg_extract.sec}) in preparation for extraction of diffuse emission (\S\ref{diffuse.sec}).
All point sources are masked in the individual, high-resolution, exposure maps; at this scale only the larger masks (on mostly off-axis sources) are visible.
\label{emap.fig}}
\end{figure}

Many \Chandra\ studies exhibit one or more of five characteristics that significantly complicate this familiar workflow.
\begin{enumerate}
\item In many studies hundreds to thousands of point sources can be readily identified; executing the workflow is intractable without significant automation.

\item Numerous sources with very few detected counts are identified in most \Chandra\ studies; common statistical methods based on large-N assumptions break down at several points in the workflow.

\item Many studies require covering a large field of view with multiple \Chandra\ pointings (e.g., Figure~\ref{emap.fig}).
For many reasons, such mosaicked pointings usually overlap significantly and/or are observed at a variety of roll angles.
Thus, many sources are observed multiple times at very different locations on the \ACIS\ detector.
Analysis of such sources can be very complex, because the \Chandra\ point spread function (PSF) exhibits large variations in size and shape across the focal plane.\footnote{See Figure~8 and Figure~10 in the \anchorparen{http://cxc.harvard.edu/cal/Hrma/users_guide/}{HRMA User's Guide}.
}
Analysis of diffuse emission is also complicated by multiple pointings, since a single diffuse region may be only partially covered by a particular \ACIS\ observation.

\item Some fields (such as deep exposures of rich star clusters, the Galactic Center, or nearby galaxies) are crowded, with adjacent PSFs in close proximity.  In such conditions, simple approaches to source extraction and background estimation are not adequate.

\item Scientifically relevant diffuse emission must often be extracted and studied while avoiding contamination from point sources (e.g., in star forming regions, the Galactic Center, and galaxy clusters in deep fields). 
\end{enumerate}

Since our \Chandra\ studies often exhibit all five of these issues, we have developed a set of data analysis methods that incorporate a variety of enhancements to standard techniques.
Publicly available software implementing these methods---the {\it ACIS Extract} package---has been cited in publications for at least 50 \Chandra\ targets. 

This paper seeks to describe these methods at a moderate level of detail.
Our primary purpose is to encourage other \ACIS\ observers to consider whether the various departures from standard techniques described here may be useful in their own projects (regardless of what tools and procedures those observers use to implement those methods).
Our secondary purpose is to document the data analysis methods used in our own \Chandra\ studies.
In many cases, the only documentation available for software and data analysis methods is on-line; thus this paper is liberally footnoted with relevant URLs.  We acknowledge that URLs are more ephemeral than journal citations but we believe that they are better than no documentation at all.  

The high-level structure of our data analysis workflow differs from standard practice in two ways.
First, for technical reasons, the optimal data cleaning steps for point sources and diffuse sources differ for \ACIS\ data (\S\ref{L1_L2.sec}); thus Figure~\ref{workflow.fig} shows separate ``cleaning'' operations on those two branches of the analysis.
Second, since the point source extraction process generates estimates of source position (\S\ref{src_positions.sec}) and source validity (\S\ref{signif.sec}) that are expected to be better than the estimates made by typical source detection procedures, we choose to adopt an iterative workflow (\S\ref{iterative_detection.sec} and dashed boxes and lines in Figure~\ref{workflow.fig}) in which the point source detection process merely nominates candidates that are then repositioned and possibly discarded as likely noise peaks after extraction results are in hand.

Most of this paper describes the individual data analysis tasks implied by Figure~\ref{workflow.fig}, emphasizing the changes to standard methods that we have adopted.
We assume the reader is familiar with \ACIS\ data and with standard analysis methods, both of which are well described by the 
\anchorfoot{http://cxc.harvard.edu/ciao/threads}{Chandra Science Threads.}
Throughout the text, we make liberal use of footnotes that direct the reader to on-line documentation, much of it provided by the Chandra X-ray Center (CXC), that is useful for understanding \ACIS\ data analysis and issues.
The process of preparing L2 data products is discussed in \S\ref{L1_L2.sec}.
Point source detection is reviewed in \S\ref{source_det.sec}.
Extraction of point sources and background estimation are presented in \S\ref{extraction.sec}.
Section~\ref{merging.sec} describes our approach to handling multiple observations of a source.
Estimation of observed and intrinsic source properties is discussed in \S\ref{src_prop.sec}.
In \S\ref{diffuse.sec} we describe modifications of point source methods for use on diffuse sources.

It is critical for the reader to recognize that the methods described here are not unique, definitive, or optimal for all purposes.   Alternative approaches to \ACIS\ data analysis are provided by the 
\anchorfoot{http://asc.harvard.edu/csc/}{Chandra Source Catalog} developed by the CXC, and by analysis tools such as \anchorfoot{http://xassist.pha.jhu.edu/zope/xassist}{{\it XAssist}} \citep{Ptak03} and \anchorfoot{http://cxc.harvard.edu/contrib/yaxx/}{{\it yaxx}} \citep{Aldcroft06}
developed by other researchers.


\section{The {\em ACIS Extract} Package}

Although the focus of this paper is to discuss data analysis techniques, rather than implementation of those techniques, in fact most of the software and recipes that we use for \ACIS\ data analysis are publicly available in a package called \anchorfoot{http://www.astro.psu.edu/xray/acis/acis_analysis.html}{{\it ACIS Extract}} (\AEacro), which was first released to the community in 2002.
\setcounter{PSUfootnotemark}{\thefootnote}
This paper will refer to \AEacro\ when discussing any data analysis method that is implemented in {\it ACIS Extract}.
We do not attempt to describe here either the full capabilities of \AEacro\ or how to use the software, but instead refer the reader to the extensive 
\AEacro\ manual and recipes.
\AEacro\ can also be applied to X-ray data from the EPIC instrument on the {\em XMM-Newton} observatory, with reduced capabilities and documentation.

\AEacro\ significantly automates the extraction and analysis of both point-like and diffuse sources; our largest project to-date \citep{Townsley10} involved 14,000 point sources and complex diffuse emission in a mosaic of 38 separate \ACIS\ observations, shown in Figure~\ref{emap.fig}.
\AEacro\ relies heavily on tools in the 
\anchorfoot{http://cxc.harvard.edu/ciao/}{{\it Chandra Interactive Analysis of Observations}} (\CIAO) package \citep{Fruscione06}.
Other software packages employed by \AEacro\ include the 
\anchorfoot{http://idlastro.gsfc.nasa.gov/}{{\it IDL Astronomy User's Library}} \citep{Landsman93}, \anchorfoot{http://space.mit.edu/ASC/MARX}{\MARX,} 
\anchorfoot{http://hea-www.harvard.edu/RD/ds9/}{{\it SAOImage DS9},}, 
\anchorfoot{http://heasarc.nasa.gov/ftools/ftools_menu.html}{{\it FTOOLS}} \citep{Blackburn95}, \anchorfoot{http://heasarc.gsfc.nasa.gov/lheasoft/xanadu/xspec/}{\XSPEC} \citep{Arnaud96}, \anchorfoot{http://www.latex-project.org/}{{\it LaTeX2e}} \citep{Lamport94}, and our own \anchor{http://www.astro.psu.edu/xray/acis/acis_analysis.html}{{\it Tools for \ACIS\ Review and Analysis}}\footnotemark[\thePSUfootnotemark] (\TARA).   
\AEacro\ is written in the \anchorfoot{http://www.ittvis.com/ProductServices/IDL.aspx}{\IDL\ language.}

Several dozen studies by several independent groups (including at least nine Large and Very Large projects) have employed \AEacro.  Examples include:
\begin{description}
\item[extragalactic survey fields:] \Chandra\ Deep Field North \citep{Alexander03};
\Chandra\ Deep Field South \citep{Luo08, Lehmer05};  
the Serendipitous Extragalactic X-ray Source Identification program \citep{Eckart06};  
SSA22 \citep{Lehmer09}

\item[lensed quasars:] PG~1115+080 \citep{Pooley06}, 1RXS~J1131-1231 \citep{Blackburne06}, survey of 10 \citep{Pooley07}

\item[galaxy clusters:] Coma Cluster \citep{Hornschemeier06}, Abell~85 and Abell~754 \citep{Sivakoff08a}, various \citep{Fassnacht08}

\item[nearby galaxies and LINERs:] IC~10 \citep{Bauer04b}, sample of LINERs \citep{Flohic06}, SN~2006gy in NGC~1260 \citep{Smith07}, M~33 \citep{Plucinsky08}, Centaurus~A \citep{Sivakoff08b}, SN~1996cr in Circinus \citep{Bauer08}, NGC~6946 and NGC~4485/4490 \citep{Fridriksson08}

\item[globular clusters:] 47~Tucanae \citep{Heinke05}, Terzan~1 \citep{Cackett06}, Terzan~5 \citep{Heinke06}, NGC~288 \citep{Kong06}, M30=NGC~7099 \citep{Lugger07}, NGC~6366 and M~55 \citep{Bassa08}, G1 \citep{Kong09}

\item[the Galactic Center:] point sources \citep{Muno03, Muno06a, Muno09} 

\item[young stellar clusters and star formation regions:]  M~17 \citep{Townsley03, Broos07}, the Orion Nebula \citep{Getman05}, L~1448 \citep{Tsujimoto05}, Cep~B \citep{Getman06}, Wd~1 \citep{Muno06b}, 30~Dor \citep{Townsley06a, Townsley06b}, W~49A \citep{Tsujimoto06}, NGC~6357 \citep{Wang07}, RCW~49 \citep{Tsujimoto07}, IC~1396N \citep{Getman07}, Coronet cluster \citep{Forbrich07}, Tr~16 \citep{AlbaceteColombo08}, W~3 \citep{Feigelson08}, CG~12 \citep{Getman08}, the Rosette Nebula \citep{Wang08,Wang09,Wang10}, NGC~6334 \citep{Feigelson09}, Cygnus~OB2 \citep{AlbaceteColombo07,Wright09}, and the Carina Nebula \citep{Townsley10}.  
\end{description}
Note that the present paper describes the capabilities of \AEacro\ as of 2009 November, and that not all features were present in earlier studies.

\section{DATA PREPARATION \label{L1_L2.sec}}

The \anchorfoot{http://asc.harvard.edu/cda/}{Chandra Data Archive} provides cleaned and calibrated X-ray data products, known as \anchor{http://cxc.harvard.edu/ciao/dictionary/levels.html}{``Level 2 Data''} (L2).
We rebuild L2 data products from the more primitive L1 products, also found in the archive, in order to apply additional processing steps.
Much of our L1-to-L2 processing \citep[Appendix~B]{Townsley03} will not be described in detail here because it follows the standard recommendations shown in the \anchorfoot{http://cxc.harvard.edu/ciao/threads}{CXC's Science Threads.}

For example, like many observers we take the precaution of verifying and improving (if possible) the astrometry of every observation, even though the absolute astrometry assigned to \Chandra\ observations using the star tracker aspect solution is often quite accurate  \anchorfoot{http://cxc.harvard.edu/cal/ASPECT/celmon/}{(${\sim}0.6$\arcsec, 90\% confidence radius}).
Astrometric alignment is particularly important when multiple observations overlap, so that the single celestial position adopted for a source will produce well-positioned extraction apertures in each of its constituent observations.
Our procedure for this task is similar to the \anchorfoot{http://cxc.harvard.edu/ciao/threads/reproject_aspect/index.html}{astrometry science thread} provided by the CXC.
We typically align each \ACIS\ observation to a published astrometric catalog,\footnote{
Galactic X-ray sources are often found in the \anchorfoot{http://www.nofs.navy.mil/nomad/}{Naval Observatory Merged Astrometric Dataset} (NOMAD) \citep{Zacharias04} or the near-infrared band \anchorfoot{http://www.ipac.caltech.edu/2mass/}{2MASS Point Source Catalog} \citep{Skrutskie06}, which are on the accurate Hipparcos reference frame.
}
rather than to a reference \ACIS\ observation, using preliminary \ACIS\ sources in the inner $8\arcmin \times 8\arcmin$ portion of the field identified by the \anchorfoot{http://asc.harvard.edu/ciao/download/doc/detect_manual}{{\it wavdetect} tool} \citep{Freeman02}.
Our current catalog matching procedure derives only offset corrections (no roll correction), which are applied to the observation's aspect file and event data using the \CIAO\ tools {\it wcs\_update} and {\it reproject\_events}.
In favorable circumstances with many matching sources, \Chandra\ fields can be aligned to the reference astrometric frame to better than $0.1$\arcsec\ precision.

Other standard L1-to-L2 processing steps are applied.  We remove events whose \anchorfoot{http://cxc.harvard.edu/ciao/dictionary/grade.html}{``grades''} are in the standard set of ``bad'' grades; we remove events arriving during time intervals designated as ``bad''; we remove events arriving during periods of \anchorfoot{http://cxc.harvard.edu/ciao/threads/acisbackground/index.py.html}{very high instrumental background} due to solar activity; we construct an exposure map for each observation.
In the early years of the \Chandra\ mission we developed and implemented a technique to mitigate the effects of charge transfer inefficiency (CTI) in both front- and back-illuminated \ACIS\ CCDs \citep[available in the Physics database of ADS]{Townsley00,Townsley02}.  Since this method has been incorporated into the standard data processing by the CXC and its calibration is maintained by them, we now use the \anchorfoot{http://cxc.harvard.edu/ciao/why/cti.html}{CXC's version of the CTI corrector,} part of the \CIAO\ tool {\em acis\_process\_events}.

A few aspects of our L1-to-L2 processing warrant discussion:
\begin{description}

\item [Improving event locations]  
Two procedures are used here.  First, standard CXC pipeline processing adds a $\pm 0.25$\arcsec\ \anchorfoot{http://cxc.harvard.edu/ciao/why/acispixrand.html}{random number to each event's position,} blurring the excellent PSF at the center of the field, as discussed in presentations at the Chandra Calibration Workshop by 
\anchorfoot{http://cxc.harvard.edu/ccr/proceedings/05_proc/presentations/marshall3}{Marshall,} 
\anchorfoot{http://cxc.harvard.edu/ccr/proceedings/05_proc/presentations/pease}{Pease (page 8),} and 
\anchorfoot{http://cxc.harvard.edu/ccr/proceedings/05_proc/presentations/smith/pg03.html}{Smith (page 3).}
We disable this randomization when the data are reprocessed by the  \CIAO\ tool {\em acis\_process\_events}.  

Second, the positions of events with non-zero event grades (i.e., where some charge appears in neighboring pixels) can be somewhat improved by ``sub-pixel resolution'' algorithms described by \citet{Tsunemi01,Mori01} and \citet{Li04}.\footnote
{
Both the \anchorparen{http://wwwxray.ess.sci.osaka-u.ac.jp/~mori/chandra/index.html}{Tsunemi} and \anchorparen{http://www.cis.rit.edu/people/faculty/kastner/SER/ser.html}{Li} groups have released tools implementing their sub-pixel resolution algorithms. 
}
Either of these two procedures will tighten the PSF in the inner portion of the field; the degree to which the PSF is improved depends upon the fraction of multi-pixel events, which in turn depends on the spectrum of the source and whether the source is imaged on a front-illuminated or a back-illuminated CCD.


\item [Bad Pixel List]
For most studies we use and recommend a less-aggressive \anchorfoot{http://cxc.harvard.edu/ciao/dictionary/bpix.html}{Bad Pixel List} than the one produced by \CIAO.  
The \ACIS\ pixels in the default list that we choose to revive\footnote{
A new observation-specific Bad Pixel Table is constructed by re-running the \CIAO\ tool {\em acis\_run\_hotpix} with its {\em badpixfile} parameter pointing to an edited list of permanent bad pixels in place of the default list.
}
were originally deprecated because they have \anchorfoot{http://cxc.harvard.edu/cal/Acis/Cal_prods/badpix/index.html}{elevated background at very low energies} (usually $<0.5$~keV, occasionally up to 1~keV) that can cause problems for the analysis of very soft diffuse structures.
Because the instrumental background increases sharply below $0.5$~keV on the front-illuminated CCDs that make up ACIS-I, event energies $<0.5$~keV are customarily ignored anyway.
We prefer to accept the small residual increased background above $0.5$~keV than to lose observatory effective area.
Our custom Bad Pixel List recovers $>4$\% of the columns in the \ACIS-I array.
A somewhat larger improvement in effective area (averaged over \ACIS-I) is achieved because standard processing of the Bad Pixel List discards events detected in the two columns adjacent to a bad column as well as the bad column itself.

\item [Bifurcated Workflow]
One component of the \ACIS\ instrumental background is a cosmic ray artifact known as  \anchorfoot{http://cxc.harvard.edu/ciao/why/afterglow.html}{``afterglow,''} operationally defined as a group of events appearing at nearly the same location on the detector in nearly consecutive CCD frames.
Eliminating afterglow events is essential to any project that seeks to detect weak point sources, since a group of afterglow events can easily be mistaken as a weak point source.
Prior to 2004 the CXC used a tool named {\em acis\_detect\_afterglow} to identify afterglow events; this aggressive tool is quite effective (few false negatives), but suffers from many false positives---events from even moderately bright sources that are mistakenly identified as afterglows.
In 2004 the CXC adopted an alternative tool named {\em acis\_run\_hotpix}; this gentle tool effectively controls false positives, but misses most afterglow series containing fewer than 10 events.
In many \ACIS\ observations, even these short afterglow series will be interpreted as statistically significant point source detections.

Two additional standard techniques for reducing the \ACIS\ instrumental background suffer the same problem of false positives near bright sources.
The first involves \anchorfoot{http://cxc.harvard.edu/ciao/threads/aciscleanvf/}{an event grading technique} (implemented by the \CIAO\ tool {\em acis\_process\_events}) that is available for data taken in Very Faint Mode.
The second involves removing a \anchorfoot{http://cxc.harvard.edu/ciao/why/destreak.html}{type of artifact found on the ACIS-S4 CCD} (implemented by the \CIAO\ tool {\em destreak}).

The lowest instrumental background level can be obtained by applying these two cleaning procedures and the aggressive afterglow detection procedure.
This aggressive cleaning scheme is preferred when searching for weak point sources and when studying diffuse emission, but is not appropriate when extracting bright sources because the number, spatial distribution, and spectral distribution of events mistakenly removed by these algorithms (false positives) near bright sources is not well known.

Thus, we recommend the bifurcated data reduction workflow shown in Figure~\ref{workflow.fig}.
An aggressively cleaned L2 event list is used to detect point sources and to extract diffuse sources, whereas point sources are extracted from an L2 event list that has been only mildly cleaned.
This strategy is not ideal because point source extractions will occasionally contain some detectable afterglow events contaminating the actual X-ray events that produced the detection.
We mitigate this problem by checking each extracted source for residual afterglow events and flagging sources dominated by afterglows.

Appendix~\ref{spare_bit.sec} shows one method for applying both afterglow detection tools and for then performing either the gentle or aggressive cleaning criteria discussed above.
Recipes and software implementing this L1-to-L2 processing are available upon request.
\end{description}


\section{POINT SOURCE DETECTION \label{source_det.sec}}

\subsection{An Iterative Source Detection Strategy \label{iterative_detection.sec}}

Typically, the process of source detection is a distinct precursor to source extraction and analysis.
Source detection algorithms must estimate some sort of significance for a putative source's signal with respect to the background expected to contaminate that signal.
In other words, source detection algorithms must perform some type of extraction of a proposed source, estimate the local background, and compute some type of significance statistic from those two quantities.

Much of this paper describes the algorithms we have developed to perform those same three tasks---extraction (\S\ref{extraction.sec}), background estimation (\S\ref{bkg_extract.sec}), and characterizing source significance(\S\ref{src_prop.sec})---on complex \ACIS\ campaigns involving multiple misaligned observations of crowded fields of point sources.
We expect these careful algorithms to produce higher quality results for point sources than can possibly be produced by typical source detection schemes that operate on binned images, do not have knowledge of the \Chandra\ PSF, and are not designed for multiple misaligned observations.

Thus, we have adopted and we recommend a source detection strategy that combines the convenience and efficiency of traditional source detection tools with the accuracy achieved only by detailed extraction and analysis of an entire catalog.
First, a liberal catalog of candidate point sources is obtained from a variety of traditional source detection methods, run with aggressive thresholds that seek to find weak sources but are expected to produce significant numbers of spurious detections.
We extract and characterize the significance of the candidates, and then prune those found not to be significant.
\citet{Nandra05} independently developed a similar strategy.

The trimmed catalog is then extracted again, re-pruned, and so on until all the remaining sources are deemed significant.
This iterative pruning procedure, depicted in Figure~\ref{workflow.fig} as a loop through the steps ``PS catalog,'' ``PS extraction,'' and ``prune/reposition sources,'' is necessary because construction of both source apertures (\S\ref{extract_reg.sec}) and background regions (\S\ref{bkg_extract.sec}) in crowded fields can be more appropriately performed when the existence of neighboring sources is taken into account.
In contrast, typical detection algorithms estimate background for a putative source using the nearby pixels in an image, without explicit regard for possible contamination by nearby sources.
Typically, only one iteration is needed for situations where source crowding is unimportant.

We test the significance of candidate sources in three energy bands (0.5--2~keV, 2--7~keV, and 0.5--7~keV) and adopt the highest value.
We prefer the upper limit of 7~keV instead of the more typical value of 8~keV because there is a line in the instrumental background spectrum at 7.47~keV.
When a source has been observed multiple times, we adopt the highest significance among all combinations of observations (\S\ref{composite_discard.sec}).

\subsection{Establishing the Candidate Source List \label{cand_src.sec}}

We construct the list of candidate sources using a combination of several methods.  
Many sources can be identified by a wavelet-based detection algorithm for Poisson images that has been widely used on \Chandra\ data--- the \CIAO\ tool \anchorfoot{http://asc.harvard.edu/ciao/download/doc/detect_manual}{{\it wavdetect}} \citep{Freeman02}---run with the liberal threshold of $1 \times 10^{-5}$ rather than the commonly used level of $1 \times 10^{-6}$ in order to nominate as many actual sources as possible without nominating an excessive number of candidates that will later be found to be insignificant.
For a simple field with one pointing, twelve images are searched---corresponding to three energy bands (0.5--2~keV, 2--7~keV, and 0.5--7~keV) and four image pixel sizes (0.25\arcsec, 0.5\arcsec, 1\arcsec, and 2\arcsec) to sample appropriately the variable \Chandra\ PSF.
All the {\it wavdetect} catalogs are merged---pairs of catalogs are matched using the algorithm described in \S\ref{matching.sec}, and the most accurate position from each matching pair is retained.
The resulting catalog is visually examined to remove any obvious duplicates remaining.

\begin{figure}[htb]
\centering
\includegraphics[width=0.5\textwidth]{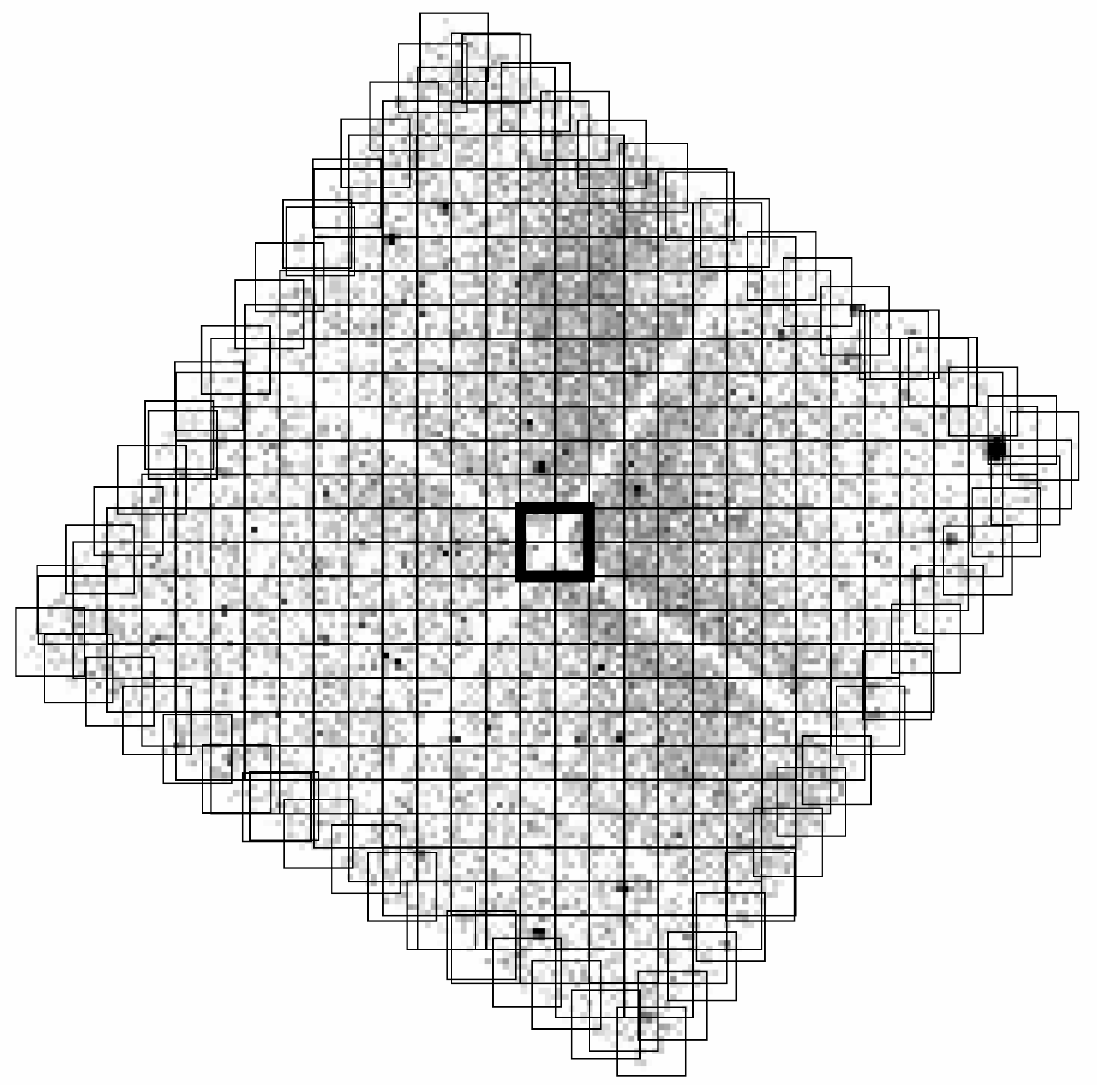}
\caption{
Many $1.5\arcmin \times 1.5\arcmin$ image reconstruction tiles covering an ACIS-I pointing.
Tiles nominally overlap by 1/4 along both axes.
Individual tiles can be seen at the field edges; a single tile at the field center is highlighted. 
Each tile image is reconstructed using a local \Chandra-\ACIS\ PSF; peaks in the reconstruction produce point source candidates.
\label{tiles.fig}}
\end{figure}

Wavelet decomposition of images is often not effective in resolving closely spaced sources, as the method is designed to find structures on a range of spatial scales. 
For such situations, image reconstruction algorithms that remove the blurring effects of the point spread function can significantly improve source identification.  
We search for faint and crowded sources by locating peaks in reconstructed images obtained with the Lucy-Richardson algorithm \citep{Lucy74}, implemented in the \anchorfoot{http://idlastro.gsfc.nasa.gov/}{{\em IDL Astronomy User's Library}.}
Similar image reconstructions have been invaluable for achieving effective spatial resolution as good as ${\sim}0.3$\arcsec\ FWHM in observations of \Chandra\ targets, for example a jet in the \Chandra\ first-light target \citep{Chartas00} and SN~1987A \citep{Burrows00}.
As the PSF varies strongly across the \Chandra\ field due to the telescope optics, the image reconstruction is performed on many small overlapping tiles (Figure~\ref{tiles.fig}) using local PSFs.  
The candidate sources identified in these tiles are merged with those obtained with the {\it wavdetect} algorithm.

\begin{figure}[htb]
\centering
\includegraphics[width=1.0\textwidth]{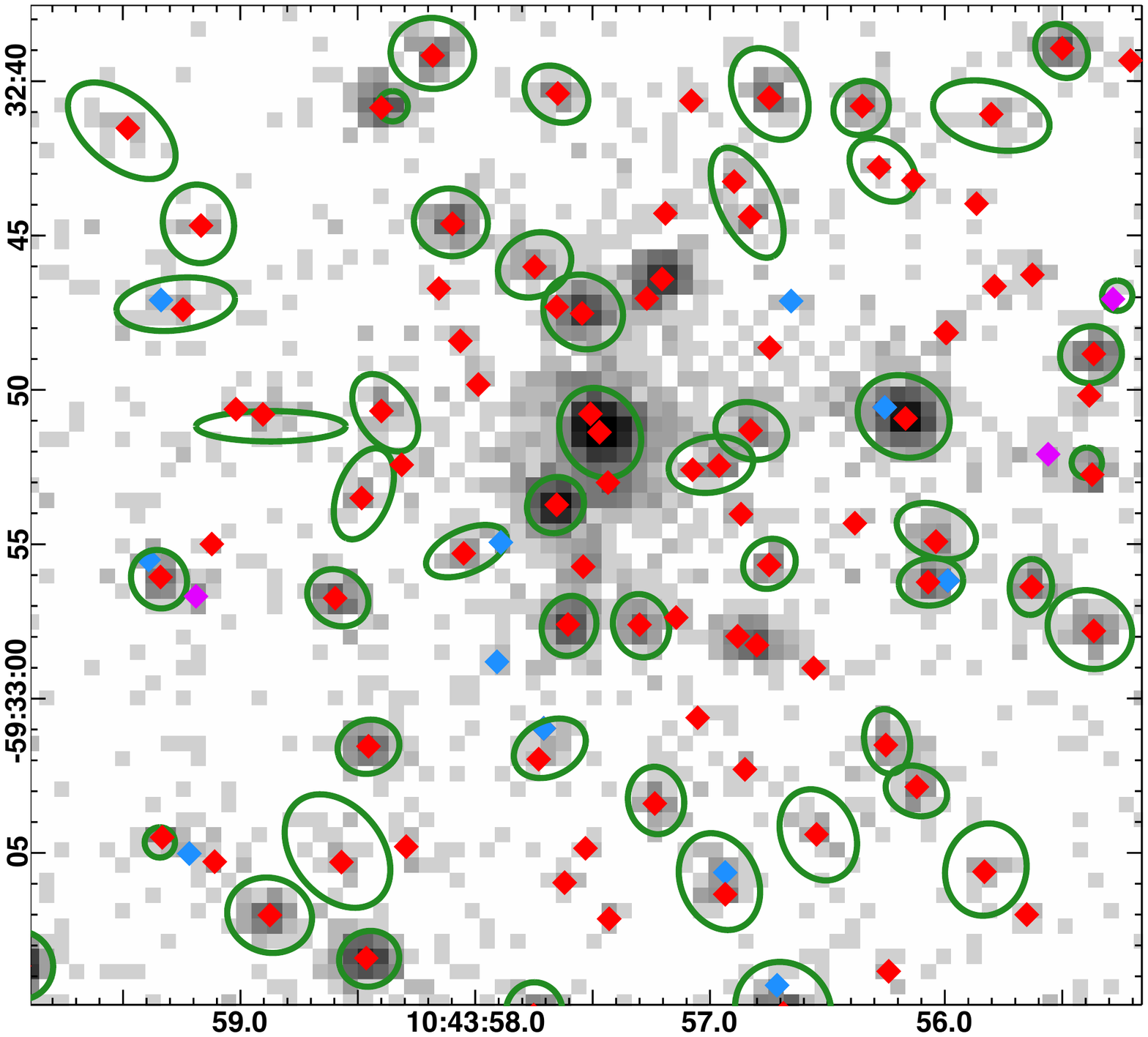}
\caption{The central 100 \ACIS\ X-ray sources (diamonds) identified in the crowded core of the young star cluster Trumpler~14 \citep{Townsley06c,Townsley10} by combining {\it wavdetect} sources (50 green ellipses) and peaks in a reconstructed image (\S\ref{cand_src.sec}).  
The underlying image shows the \ACIS\ data binned at 0.5\arcsec\ per pixel; the coordinate axes are J2000 Right Ascension and Declination.  
X-ray source extraction apertures constructed by \AEacro\ (\S\ref{extract_reg.sec}) are not shown.
The 86 sources confirmed by near-infrared observations using NTT/SOFI and VLT/NACO \citep{Ascenso07} are shown in red; the median offset between X-ray and IR positions is 0.14\arcsec.
An additional 3 sources (magenta) are confirmed by new VLT/HAWK-I observations (Thomas Preibisch, private communication).  
The 11 sources currently not identified in other wavebands are shown in blue.
\label{Tr14.fig}}
\end{figure}

Figure~\ref{Tr14.fig} shows an example of the effectiveness of image reconstruction in regions with crowded, faint sources.  
The {\it wavdetect} procedures located 50 sources in this sub-image of the center of the young stellar cluster Trumpler~14; an additional 50 are found as peaks in the reconstructed images \citep{Townsley06c,Townsley10}.  
The reliability of most of these sources is confirmed; 89 of the 100 reconstruction sources coincide with stars detected in deep, high-resolution near-infrared exposures \citep[Thomas Preibisch, private communication;][]{Ascenso07}, some with separations $<$1\arcsec.  
Of the 11 unconfirmed X-ray sources, 9 appear in close pairs where the other member of the pair is confirmed by an IR counterpart.  Testing the validity of these X-ray sources will require very high-resolution, sensitive IR data, as it may be difficult to find faint, close companions in IR observations of such crowded regions suffused by diffuse IR emission.  Since the X-ray flux of a young star does not correlate closely with its IR brightness, it is not unreasonable for us to find close X-ray pairs that are not seen in IR images, and vice versa.

When a project's scientific goals include the search for faint X-ray emission from a previously-defined catalog of interesting objects found in other wavebands, the candidate X-ray source catalog can be supplemented with source positions that are not derived from the \ACIS\ image.
The \AEacro\ procedures can then judge whether significant emission is present from each single source, and the positions can be `stacked' for high-sensitivity examination of the collective properties of the user-provided catalog (\S\ref{stack.sec}).

Finally, we often supplement the candidate source catalog with the coordinates of suspected point sources identified from visual review (e.g., with the \DSnine\ visualization program) of the X-ray data.
This visual review also allows removal of obviously spurious candidate sources that occasionally emerge (e.g., from detector artifacts such as bright source readout-streaks) and duplicate sources.

\subsection{Thresholding a Source Significance Statistic \label{signif.sec}}

In typical source detection schemes, the existence of a candidate source is evaluated using the signal-to-noise ratio (SNR), which is a photometry value divided by its uncertainty.
When photometry values have Gaussian distributions, then a SNR threshold directly corresponds to a level of significance in a statistical test of the null hypothesis that there is no source signal present, only background.

However, most \ACIS\ sources are quite weak with very few counts extracted from the source aperture, and in crowded fields sometimes very few counts available to estimate the background.
Gaussian approximations to photometric confidence intervals in such cases can be quite poor.
We prefer to test directly the null hypothesis that a candidate source does not exist using the method described by \citet[][Appendix~A2]{Weisskopf07} based on the Poisson distribution.
Computation of this significance statistic, $P_B$, is described in Appendix~\ref{signif.app}.

The observer must set, either {\it a priori} or after careful examination of the image and possible multi-wavelength counterparts, a threshold value of $P_B$ that strikes a reasonable balance between the competing goals of low false detection rates and high sensitivity.
Analytical methods to estimate false detection rates and sensitivity have been developed \citep[e.g., by][]{Nandra05,Georgakakis08}, however extending these methods to studies that involve crowded source apertures (\S\ref{extract_reg.sec}) and multiple overlapping pointings (\S\ref{composite_discard.sec}) is not straightforward. 
Monte Carlo simulation of the detection process \citep[e.g., by][]{Cappelluti07,Kim07} is a powerful technique to study false detection rates and sensitivity, however simulation of our complex detection process would require unreasonable quantities of both human and computing resources.
Thus, we do not currently have an objective or authoritative way to set detection thresholds.  
As with most source detection tasks, the scientist's subjective judgment is critical for setting criteria for source existence, and high-quality observations in other wavebands are invaluable for informing that judgment.

\section{POINT SOURCE EXTRACTION   \label{extraction.sec}}

The process of ``extracting'' a putative point source consists of several tasks: defining an appropriate aperture around the source position, defining an appropriate background region that is expected to be a good estimator for the background contaminating the source aperture, collecting the observed events within the aperture and background regions, and constructing a model for the response of the observatory that correctly calibrates the extraction so that intrinsic source properties can be derived.

When the source is assumed to be point-like, all of these tasks are best performed by algorithms that consider the shape of the local \Chandra-\ACIS\ point spread function (PSF).
Thus, \AEacro\ begins the extraction process by constructing a model of the local PSF (Appendix~\ref{PSFs.sec}).

\subsection{Construction of Extraction Aperture and Mask Region\label{extract_reg.sec}}

Many \Chandra\ observers obtain extraction apertures for their point sources either from the \CIAO\ tool
\anchorfoot{http://asc.harvard.edu/ciao/download/doc/detect_manual}{{\it wavdetect}} \citep{Freeman02} (the most popular source detection tool in the \Chandra\ community), or by drawing the apertures by eye.\footnote
{
See for example the analysis threads \anchorparen{http://asc.harvard.edu/ciao/threads/psextract/}{``Using psextract to Extract ACIS Spectra and Response Files for Pointlike Sources''} and \anchorparen{http://asc.harvard.edu/ciao/threads/pieces/}{``Step-by-Step Guide to Creating ACIS Spectra for Pointlike Sources''}.
}
Because {\it wavdetect} does not have explicit knowledge of the \Chandra\ PSF, and because it seeks to detect both point-like and extended sources, it sometimes produces aperture shapes that are radically different from the PSF (e.g., ellipses with extreme eccentricity).
An additional concern with apertures from {\it wavdetect} or visual inspection is the risk of introducing an upward bias into photometry because both tend to ``follow the light''---including only regions of the image where, by chance, the source produced an excess of events over the background.
In our opinion, a region derived from the local PSF is the most objective and appropriate extraction aperture for point sources.

When a source is ``extended'' (resolved by \Chandra) the spatial distribution of its observed events do not follow the local PSF, and application of the point source extraction techniques in this section (\S \ref{extraction.sec}) will underestimate the flux of the source.
Such sources are more properly designated as ``diffuse'' and should be extracted by procedures that do not make the point-like assumption (see \S \ref{diffuse.sec}).
No automatic procedure for defining the extraction aperture of such a source seems feasible---one must either define an aperture that follows the observed light (as {\it wavdetect} does), or apply {\it a priori} information about the structure of the source to define a suitable aperture.
Observers often struggle to decide if a source is extended.
The \Chandra\ Source Catalog provides a sophisticated analysis of \anchorfoot{http://cxc.harvard.edu/csc/columns/srcextent.html}{source extent}, and the CXC has recently introduced a \anchorfoot{http://cxc.harvard.edu/ciao/threads/srcextent/}{science thread} for measuring source extent.
\AEacro\ does not currently address this task.

\begin{figure}[htb]
\centering
\includegraphics[width=0.5\textwidth]{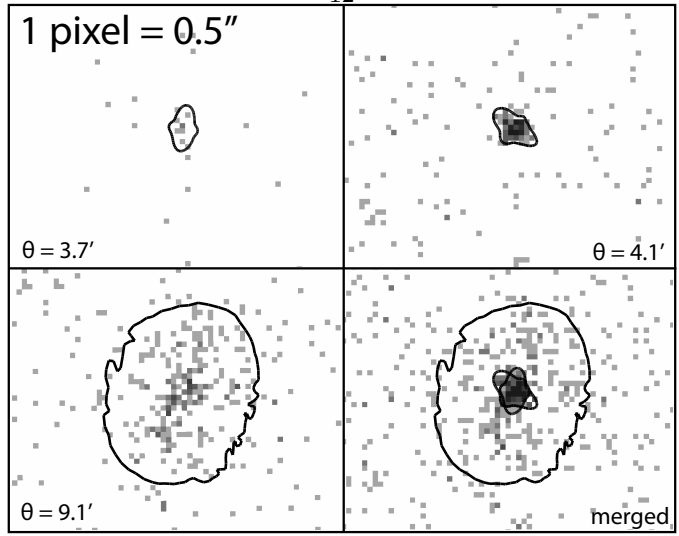}
\caption{Example of extraction apertures (contours of the local PSF) enclosing 90\% of the PSF power constructed by \AEacro\ (\S\ref{extract_reg.sec}) for a source observed at three off-axis angles---3.7\arcmin\ (upper-left), 4.1\arcmin\ (upper-right), and 9.1\arcmin\ (lower-left).
The shape and overall size of the \Chandra\ PSF varies significantly across the field of view.
Combining {\em all} of the available extractions of a source (lower-right) will in some cases produce lower-quality estimates of source properties than could be obtained by ignoring some observations (e.g., the far off-axis data in the lower-left panel).
Techniques for deciding which observations to ignore are discussed in \S\ref{composite_discard.sec}.
\label{PSF_polygons.fig}}
\end{figure}

\begin{figure}[htb]
\centering
\includegraphics[width=0.5\textwidth]{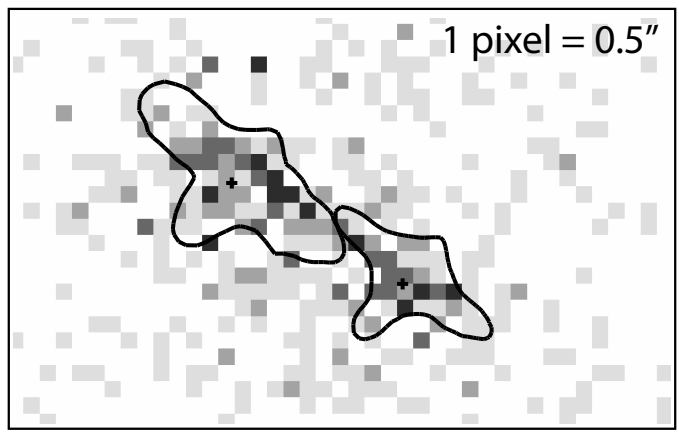}
\caption{Example of crowded extraction apertures (contours of the asymmetric local PSF at ${\sim8}$\arcmin\ off-axis) that have been reduced in size (enclosing 56\% and 44\% of the PSF power) until they do not overlap.  
These two X-ray source positions (plusses) are confirmed by a high-resolution infrared observation.
\label{reduced_apertures.fig}}
\end{figure}

Several observers employ elliptical approximations to the PSF as extraction apertures, e.g., \cite{Nandra05} and the \Chandra\ Source Catalog\footnote{
The \Chandra\ Source Catalog performs \anchorparen{http://asc.harvard.edu/csc/columns/fluxes.html}{two extractions for each source}---one using a {\em wavdetect} ellipse and one using a PSF ellipse.
}
use ellipses that enclose 70\% and 90\% of the PSF power respectively.
\AEacro\ extraction apertures are built from contours of the local PSF at 1.5~keV, as shown in Figures~\ref{PSF_polygons.fig} and \ref{reduced_apertures.fig}.
By default, \AEacro\ apertures enclose ${\sim}90$\% of the PSF power, which we have found is a reasonable trade-off between maximizing the source's signal and minimizing the background's signal in typical fields.
\AEacro\ does not currently attempt to optimize the aperture size based on the source and background levels (e.g., use a large aperture for bright sources and a small one for sources just above the background).

For many \Chandra\ observations, some of these default apertures will overlap significantly due to source crowding.
\AEacro\ iteratively reduces the aperture sizes of crowded sources until the apertures no longer overlap (Figure~\ref{reduced_apertures.fig}).
The brighter member of a pair of crowded sources maintains its default aperture until the aperture for the weaker member has been driven down to a minimum allowed size (enclosing ${\sim}40$\% of the PSF power).
Of course, some light from the wings of nearby sources may contaminate even a reduced source aperture; this light constitutes an additional background component for the source.
\AEacro\ provides a sophisticated background algorithm that models and subtracts this component (\S\ref{bkg_extract.sec}).

In a particular observation, a source may be so highly crowded that assigning minimal apertures to it and its neighbor cannot prevent overlap.
When multiple observations of the source are available, such overlapping extractions are discarded by \AEacro\ (\S\ref{composite_discard.sec}) in recognition that there are limits to our ability to estimate backgrounds (\S\ref{bkg_extract.sec}) under conditions of extreme crowding.
If all extractions of a pair of candidate sources are overlapping, then our policy is to eliminate the weaker candidate in recognition that either the detection process has mistakenly split a single physical source into two candidates, or that two physical sources are not resolved by our observation (i.e., meaningful source properties cannot be determined for each source).

A PSF model and an extraction region must be constructed independently for each observation of a source.
A pair of sources may be well-separated with nominally-sized apertures in an on-axis observation, but may suffer severe crowding with reduced apertures in an off-axis observation.
Even a source observed twice at similar off-axis angles may have apertures with very different shapes, since the azimuthal shape of the \Chandra\ PSF varies across the focal plane.\footnote{See Figure~10 in the \anchorparen{http://cxc.harvard.edu/cal/Hrma/users_guide/}{HRMA User's Guide}.
}

\subsection{Event Extraction and Calibration Data Products \label{OGIP_files.sec}}

Once apertures are determined for each observation of a source, a mostly routine extraction process is followed for each observation using standard \CIAO\ tools---{\em dmextract} selects the events within the aperture and constructs a Type I                                        
\anchorfoot{http://heasarc.gsfc.nasa.gov/docs/xanadu/xspec/fits/fitsfiles.html}{HEASARC/OGIP-compatible}
``source'' spectrum; {\em mkarf} queries the \Chandra\ Calibration Database and constructs an ancillary reference file (ARF); {\em mkacisrmf} queries the \Chandra\ Calibration Database and constructs a response matrix file (RMF). 
When an aperture dithers across multiple CCDs, standard extraction methods\footnote
{
See for example the analysis threads \anchorparen{http://asc.harvard.edu/ciao/threads/psextract/}{``Using psextract to Extract ACIS Spectra and Response Files for Pointlike Sources''} and \anchorparen{http://asc.harvard.edu/ciao/threads/pieces/}{``Step-by-Step Guide to Creating ACIS Spectra for Pointlike Sources''}.
} 
will over-estimate the source flux (by as much as 50\%) because the response at the source position of only one CCD is computed (using a single call to {\em mkarf}).
\AEacro\ avoids this mis-calibration of the extraction by computing and summing the response at the source position of all the CCDs (using multiple calls to {\em mkarf}).

\subsection{Aperture Correction \label{aperturecorr.sec}}

Perhaps the most important calibration facilitated by the PSF model is accounting for the point source light falling outside the aperture and not extracted---a standard part of optical and infrared data analysis commonly referred to as ``aperture correction.''
Such correction is necessary because the Chandra Calibration Database effective area data assume an infinitely large detector and extraction aperture.\footnote
{
For example, the {\em ahelp} page for {\em mkarf} says ``The ARF is computed assuming that the spectral extraction region is large enough to include the entire PSF (e.g., PSF fraction=1.0).''
}

\begin{figure}[htb]
\centering
\includegraphics[width=0.5\textwidth]{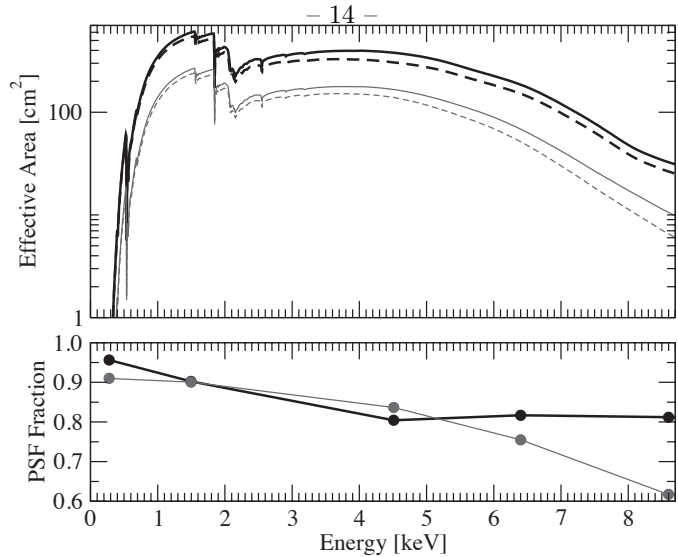}
\caption{Effective area as a function of X-ray energy (upper panel) for typical on-axis (black) and off-axis (gray) sources extracted by \AEacro\ using default extraction apertures (90\% PSF fraction at 1.5~keV, \S\ref{extract_reg.sec}).
Each aperture-corrected (dashed) curve is the product of the nominal curve (solid, from {\em mkarf}) and a PSF fraction curve (lower panel) obtained by interpolating measurements at five energies (dots).
\label{aperture_corr.fig}}
\end{figure}

Since the size of the \Chandra\ PSF varies significantly with photon energy,\footnote
{See Figures~5 and 6 in the \anchorparen{http://cxc.harvard.edu/cal/Hrma/users_guide/}{HRMA User's Guide}.
}
the aperture correction is a function of energy.
\AEacro\ builds images of the local PSF at five monochromatic energies (Appendix~\ref{PSFs.sec}), and calculates for each PSF image the fraction of the power that falls within the aperture.
Those five ``PSF fractions'' are interpolated to estimate a PSF fraction at every energy.
The product of this PSF fraction curve and the nominal ARF produced by \CIAO\ represents the true observatory effective area that is supplying photons within the aperture; this corrected ARF is carried forward in the analysis.
The energy and off-axis dependence of this aperture correction is illustrated in Figure~\ref{aperture_corr.fig}.

Omitting an aperture correction has two scientific effects.
Firstly, flux estimates are biased downward (by ${\sim}10$\% for the examples in Figure~\ref{aperture_corr.fig} for a typical observed spectrum peaking near 1.5~keV).
Secondly, the inferred {\it shape} of the source spectrum is somewhat distorted, especially for off-axis sources, because the PSF fraction varies with energy (Figure~\ref{aperture_corr.fig}, lower panel).


\subsection{Background Extraction \label{bkg_extract.sec}}

Assessment of the detection significance and the spectral properties of a weak source depend critically on estimation of unbiased background spectra for each extraction (each observation) of the source.
Often, background estimates must be performed ``locally'' for each source to account for spatial variations due to diffuse emission, or wings of nearby point sources. 
\AEacro\ supports three methods of constructing local background spectra.

\begin{figure}[htb]
\centering
\includegraphics[width=0.50\textwidth]{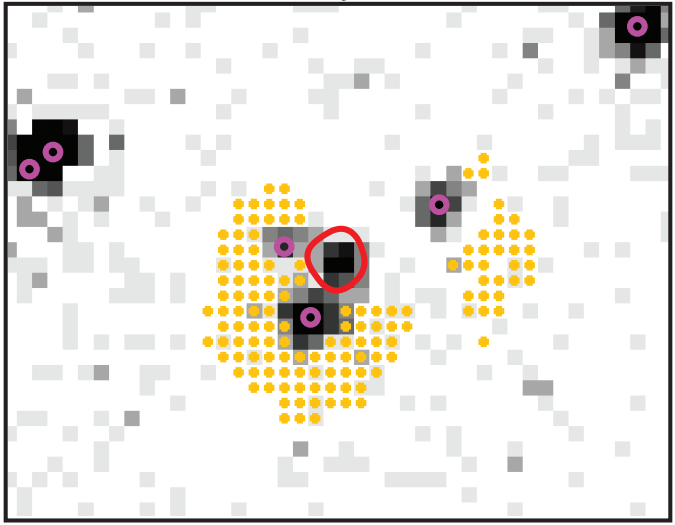}
\caption{A background region constructed by \AEacro\ (pixels marked in yellow) for a crowded source (red polygon).
The region seeks to sample neighboring sources (purple circles) in proportion to their expected contamination of the red source's extraction aperture.  
\label{better_bkgds.fig}}
\end{figure}

\begin{enumerate}
\item When all sources are well separated, a traditional approach using annular background regions is adequate.
\AEacro\ implements this approach by first masking (removing) virtually all the point source light from the data set using circular mask regions with a nominal radius that is 1.1 times a radius that encloses 99\% of the PSF.
Circular background regions are then defined independently for each source to encompass a number of background events specified by the observer.

\item \AEacro\ also provides an alternative masking algorithm that first models the surface brightness expected from the point sources (using the PSF models and rough estimates of source fluxes), and then iteratively seeks to mask (remove) all portions of the field where the ratio of that point source glow to the local background level exceeds a threshold.
This algorithm allows the large wings of bright sources to be removed with large masks, while avoiding excessive loss of background area around weak sources, as shown in Figure~\ref{emap.fig}.
Circular background regions are then defined independently for each source to encompass a number of background events specified by the observer.

\item When sources are crowded, the concept of creating background spectra from a data set cleaned of all point source light is not appropriate.
By definition, a significant component of the background for a crowded source arises from the wings of its neighbors; those wings must be modeled.

The ultimate analysis strategy would conceptually separate a source's background into a point source wing component and a traditional smooth component not associated with detected point sources.
The latter would be estimated via the masking techniques described above, and the former would be {\em modeled} via a complex multi-source spatio-spectral modeling process. 
Lacking the resources to implement that strategy, we have implemented a less complex strategy that seeks to estimate both components within each source's aperture by carefully {\em sampling} the observed event data. 

\AEacro\ first constructs a spatial model for every source (using the PSF models and rough estimates of source fluxes), and then iteratively constructs a background region for each source that seeks to sample fairly the light that each neighbor's wing is depositing into the source aperture.
In the simple case where the source has one neighbor that is contributing background to the source aperture, the algorithm will tend to build a background region that is an annulus around the neighbor, as shown in Figure~\ref{better_bkgds.fig}.
As the background region grows during the iteration, it will seek to avoid the wings of more distant sources that are not contributing light to the source aperture; thus for uncrowded sources this algorithm produces similar background regions as algorithm \#2.
\AEacro\ tries to balance the competing goals of acquiring the desired number of background events, fairly representing each background component (wings from each neighboring point source, diffuse emission, and instrumental background), and sampling the background locally (acquiring background events in a compact region around the target source).
\end{enumerate}

For either masking-based approach, the observer can manually specify masks to remove portions of the field that are not suitable background estimators, such as CCD readout streaks associated with bright sources.  
For the spatial modeling-based approach, the observer can supply a spatial model for background structures that are not point sources, such as CCD readout streaks; such structures will be avoided or included in individual background regions to the extent that the source aperture is contaminated by them.

Regardless of how the background region is defined, we believe that the scaling traditionally applied to the background spectrum---the ratio of the geometric areas of the source aperture and background region---is not appropriate for \ACIS\ data because the \ACIS\ exposure map contains significant small-scale features caused by the dithering of CCD edges and bad columns.
For example, a source lying in a so-called chip gap may have an effective exposure time that is less than half of nominal, while the majority of its background region may have nominal exposure.
\AEacro\ accounts for these exposure map features by adopting a background scaling equal to the ratio of the integrals of the exposure map over the source aperture and background region.

\section{MERGING EXTRACTIONS FROM MULTIPLE OBSERVATIONS \label{merging.sec}}

Multiple observations of the same or overlapping fields have occurred throughout the \Chandra\ mission; evolving constraints on mission operations now commonly require projects to be split into short segments. 
When the aim points and roll angles are nearly identical, it is appropriate to join the segments and treat them as a single observation \citep[e.g.,][]{Getman05}.  
But when the observations are misaligned, extraction must be performed separately on each observation.\footnote
{
The \CIAO\ thread \anchorparen{http://cxc.harvard.edu/ciao/threads/combine/}{Merging Data from Multiple Imaging Observations} states:
``The merged event list should not be used for spectral analysis, since it does not contain sufficient information to generate correct response files. The recommended technique for the spectral analysis case is to generate separate PHA, RMF, and ARF files for each observation ...''.
}

\subsection{Composite Data Products}  

In principle, X-ray properties for a source could be estimated directly from the multiple extractions of the source (multiple source spectra, background spectra, ARFs, and RMFs).
For example, photometric quantities could be estimated by maximizing a likelihood function involving terms for all of the observations (source and background extractions).
Similarly, a spectral model for the source could be derived by simultaneously fitting all of the extractions.
However, this approach is impractical for the majority of faint \Chandra\ sources, where a few counts are spread over a number of separate observations.

\AEacro\ adopts an alternate strategy---multiple extractions of a source are merged into ``composite'' data products (PSF model, source spectrum, background spectrum, ARF, and RMF), which are carried forward for analysis.
Source spectra and exposure times are summed.
Weighted averages of the PSFs, ARFs, and RMFs are computed using \IDL\ and the FTOOLS {\it addarf} and {\it addrmf}.
In order to support ``grouping'' of low-count spectra at later stages of the analysis, the composite background spectrum is represented in the same way as the source spectrum, as an integer ``counts'' vector (the sum of the extracted background spectra) with an associated  scaling value (to account for the different sizes of the source and background regions), rather than as a real-valued ``rate'' vector with uncertainties on each rate value.\footnote
{
When a fitting package groups several spectral channels together, if the spectrum is expressed as integer counts then the Poisson uncertainty on the group can be estimated accurately from the total number of counts in the group (e.g., via $\sqrt{N}$).
In contrast, grouping a spectrum expressed as a rate vector with uncertainties requires Gaussian propagation of those uncertainties; many channels in typical \ACIS\ spectra have zero or one count, with virtually meaningless uncertainties.
}
Extraction data products from each observation are retained so that the observer can analyze them separately (or simultaneously) if desired.

An important constraint on the design of the individual extractions arises from the decision to sum the integer-valued background spectra, namely that all the individual background regions must be designed to have similar scaling values.
This constraint arises because each individual background spectrum must be fairly represented in the composite background in order for Poisson ($\sqrt{N}$) uncertainty estimates to apply to the composite background.
This can be understood by considering an extreme example.
Imagine a background channel in which observation \#1 produced 25 counts with a scaling of 25 and observation \#2 produced 10,000 counts with a scaling of 10,000.
The scaled composite background rate would be $(25/25) + (1 \times 10^4/1 \times 10^4) = 2.0$.
However, the Poisson uncertainty estimated (by a fitting package) for that background rate would be far too small, since it is based on the notion that 10,025 counts were observed.
In fact, the 25-count extraction totally dominates the uncertainty on the scaled composite background, which actually has an approximate uncertainty of $\sqrt{(\frac{\sqrt{25}}{25})^2 + (\frac{\sqrt{10,000}}{10,000})^2} \simeq 0.2 $.
\AEacro\ includes code and workflow recommendations to ensure that all extractions of a source result in similar scaling of their background spectra.

\subsection{Discarding Observations \label{composite_discard.sec}}
When multiple observations of a field are available, astronomers commonly choose to disregard observations that are judged to be unhelpful to whatever astrophysical measurement is desired.
For example, in ground-based optical studies, exposures with unusually bad seeing or unusually high background might be ignored.
In all \ACIS\ studies, observers are encouraged to discard data obtained during periods of \anchorfoot{http://cxc.harvard.edu/ciao/threads/acisbackground/index.py.html}{very high instrumental background} (due to solar activity).\footnote
{
For convenience, this is commonly performed in the early stages of data analysis, guided by the damage that the enhanced background will do to the sources most susceptible to background (e.g., diffuse sources).
A somewhat better strategy would be to choose high-background periods to discard on a source-by-source basis, since very bright point sources would benefit more from extra integration time than from a reduction to their already insignificant background. 
\AEacro\ has not yet adopted this optimum strategy because implementation is difficult.  
}
In \ACIS\ studies involving multiple misaligned pointings, observers often choose to ignore data taken far off-axis (where the \Chandra\ PSF is significantly degraded\footnote
{See Figure~8 in the \anchorparen{http://cxc.harvard.edu/cal/Hrma/users_guide/}{HRMA User's Guide}.
}) 
\setcounter{HRMAfootnotemark}{\thefootnote}
if on-axis coverage is available \citep[e.g.,][]{Plucinsky08}.
\ACIS\ observers also commonly choose to search for sources both within each observation separately and within observer-defined combinations of observations \citep[e.g.,][]{Muno09}.

\AEacro\ tries to formalize this common practice of ignoring unhelpful observations by implementing three data-selection strategies that are applied independently to each source.
\begin{enumerate}

\item When the observer is interested in the {\em validity} of proposed sources using \AEacro's $P_B$ statistic (\S\ref{signif.sec}), then we recommend an \AEacro\ option that selects whatever subset of the available extractions that optimizes (minimizes) $P_B$ for each source.
Under this option, an extraction will be included only when the particular signal and background it contributes lowers $P_B$.
This option is appropriate when the observer wishes to adopt the scientific policy that a source is deemed to exist if it is significant in any observation, or in any combination of observations.
For highly variable astrophysical sources, such as young stars, this is a reasonable strategy even when multiple observations have similar PSFs.
The price paid for increased sensitivity to variable sources is an increased false detection rate from the additional number of random ``trials'' that can produce spurious detections. 

\item When the observer is interested in the {\em position} of sources, then we recommend an \AEacro\ option that selects whatever subset of the available extractions optimizes (minimizes) the expected position uncertainty (\S\ref{src_positions.sec}) for each source.
An extraction much farther off-axis than its peers will be included only when the disadvantage of its larger PSF\footnotemark[\theHRMAfootnotemark]
is outweighed by the advantage of averaging over more counts.
Since we register each of our observation to an absolute astrometric reference frame, all observations of a source are well-aligned and it is appropriate to estimate the position using only the best data we have.

\item When the observer is interested in {\em time-averaged photometric properties} (e.g., fluxes, spectra) then selecting the extractions to merge becomes more problematic.
Anytime extractions are discarded in order to optimize a photometric quantity (e.g., SNR) a bias can be introduced into all photometric properties because the discarded extraction may have, by chance, lower observed flux than the long-term average.
Thus, the observer must balance two undesirable outcomes: sources whose photometric accuracy is degraded by including very poor-quality extractions, and sources whose photometry suffers the suspicion of bias because some extractions were discarded after examining their data.
\AEacro\ offers an option that strikes this balance by discarding extractions only when retaining them would drive the SNR of the merged data set significantly below the optimal SNR.
The observer specifies the minimum acceptable ratio between the SNR achieved by the merge and the optimal SNR achievable by discarding more extractions.
This strategy tolerates limited deterioration to the SNR in order to avoid photometric bias arising from data selection.
Bright sources will tend to incorporate all observations since background is unimportant, whereas weak sources will tend to reject observations that have backgrounds much higher than their peers (e.g., those with much larger apertures). 

\end{enumerate}

\section{SOURCE PROPERTIES \label{src_prop.sec}}

After merging the appropriate set of extractions (observations) for each source, \AEacro\ calculates various photometric quantities, estimates various source properties, fits astrophysical models to source spectra using \XSPEC, and constructs light curves. 
These analysis capabilities are described in the \AEacro\ manual, but a few warrant discussion here.

\subsection{Source Position \label{src_positions.sec}}

While initial source positions are provided by a source detection process, these may not be optimal.  \AEacro\ provides three procedures for improving source positions.  One position estimate is constructed by calculating the centroid of the extracted events.  
A second estimate is determined from the peak of the spatial correlation between the \Chandra-\ACIS\ PSF (Appendix~\ref{PSFs.sec}) and the events in a neighborhood around the source.   
A third position estimate is obtained from the location of the peak in a maximum-likelihood \citep{Lucy74} image reconstruction of the source neighborhood.  
In a multi-observation reduction, the correlation and reconstruction operations are performed using a composite (multi-observation) event image and a composite PSF image because a given source may be undetected in an individual observation. 

In our experience, all three methods give very similar positions for most on-axis sources; the centroid position is simplest to calculate and is often used.  The PSF correlation position is best for sources far off-axis ($\theta \ga 5$\arcmin);  centroid positions are biased estimates of the true locations due to the asymmetry of the off-axis PSFs.  For very closely-spaced sources with overlapping PSFs (Figure~\ref{reduced_apertures.fig}), the best positions are provided by the maximum-likelihood reconstructed images.   
Since the centroid position is calculated using the extracted events, it should be re-calculated after a source is repositioned to verify convergence.

Positional errors are estimated separately along the right ascension and declination axes according to 
\begin{eqnarray}
\sigma_{RA} &=& \sigma_{model,RA}/\sqrt{N}   \\
\sigma_{Dec} &=& \sigma_{model,Dec}/\sqrt{N}   \\
\sigma_{r} &=& \sqrt{\sigma_{RA}^2 + \sigma_{Dec}^2}   \label{pos_error.eqn}
\end{eqnarray}
where $\sigma_{model,RA}$ and $\sigma_{model,Dec}$ are the standard deviations along the right ascension and declination axes of a model of the counts falling within the extraction region.
That model consists of the PSF projected onto each axis plus a flat background component appropriately scaled, both truncated by the source's extraction region; the spatial distribution of the {\em observed} data is not considered.\footnote
{
The \Chandra\ Source Catalog takes a different approach to \anchorparen{http://asc.harvard.edu/csc/columns/positions.html}{positional errors,} adopting \anchorparen{http://asc.harvard.edu/csc/why/err_ellipse_msc.html}{error ellipses} directly from the {\it wavdetect} program for single observations, and combining ellipses for multiple observations.
} 
$N$ is the total number of extracted counts.  The quantity $\sigma_r$, commonly known as the distance root-mean-square error, is generally reported as the ``1$\sigma$'' error, or 68\% confidence interval, for the source position.   


\subsection{Grouped Spectra \label{grouping.sec}}

Standard \ACIS\ spectral files divide the energy range covered by the instrument into many hundreds of pulse-invariant (PI) energy channels, many of which have very few counts for typical sources.  Energy channels are therefore frequently grouped.
\AEacro\ offers only one of the several grouping criteria found in standard grouping tools---groups are constructed to have similar signal-to-noise ratios---but improves on standard implementations in several important ways:
\begin{enumerate}

\item Standard algorithms provide no convenient way for the observer to define precisely the energy range over which spectral models will be fit.
\AEacro\ allows the observer to specify the boundaries of the first and last groups. 
For example, if the energy range 0.5--8.0~keV is specified, then the first group will span all the channels below 0.5~keV and the last group will span all the channels above 8.0~keV; these terminal groups can be subsequently ``ignored'' within a fitting package.  

\item The criterion \AEacro\ uses to define a group is a target SNR for the {\em net counts} in the group.
Standard grouping tools do not consider the background spectrum, and thus produce background-subtracted grouped spectra with SNR values lower than the target.

\item Standard algorithms are asymmetric, filling groups from low to high energies.  
For sources with fewer than several hundred counts, groups often begin with a string of empty channels and end on a channel that happens to have an observed event.  
This permits anomalies in the grouped spectra, such as narrow-width groups with overestimated flux adjacent to wide  groups with underestimated flux.
These distortions can bias the spectral fitting and inflate the best-fit $\chi^2$ value.  
\AEacro's grouping algorithm attempts to mitigate this problem by selecting group boundaries mid-way in the run of empty channels (if any) that lie between events belonging to adjacent groups. 

\end{enumerate}

\subsection{Non-parametric Characterization of Spectral Shape \label{Emed.sec}}

X-ray spectra observed with \ACIS\ energy resolution typically show some combination of continuum components, line emission, and absorption features from interstellar matter.  Accurate characterization requires fitting with non-linear astrophysical models (\S\ref{XSPEC.sec}).  However, this fitting procedure does not give meaningful spectral parameters for faint sources when a wide range of models are statistically consistent with the sparse data set.   Non-parametric estimators of spectral shape---statistics that describe the shape of the observed spectrum without assuming any astrophysical model---are commonly used for weak sources.

The oldest and most widely used statistic characterizing the shapes of X-ray spectra is the ``hardness ratio'' involving various ratios of soft and hard band counts.  These estimators (known as `Poisson proportions' in the statistical literature) 
are mathematically unstable for sources with extreme spectra where the numerator or denominator has few or zero counts \citep{Brown01}.  
\citet{Hong04} demonstrated that quartiles (three values of a random variable that divide a population into four equal  groups) of the observed event energies are superior characterizations of spectral shape for low-count observations, avoiding the problem of empty or nearly empty pre-defined energy bands.
Quartile-quartile plots can be mapped to astrophysical parameters for specified simple model families.

\begin{figure}[htb]
\centering
\includegraphics[width=0.5\textwidth]{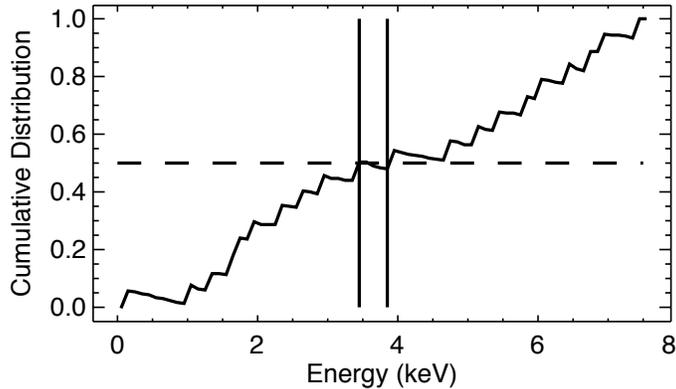}
\caption{Example calculation of the background-corrected median energy statistic.
The 20 large upward jumps in the cumulative distribution of net counts (rising curve) represent the 20 counts observed in the source aperture; the 100 small downward jumps (not individually visible) represent the 100 counts observed in the background region, scaled down to match the source aperture size.
The lowest energy (left vertical line) and highest energy (right vertical line) at which the 50th percentile (dotted line) is reached are averaged to produce an estimate of the median energy of the parent source.
Here, the median energy is ${\sim}3.6$~keV.
\label{Emedian.fig}}
\end{figure}

\AEacro\ provides estimates of the 25\% quartile, median (50\% quartile), and 75\% quartile of the observed event energies over a variety of bands.
These statistics are background-corrected, meaning that they seek to characterize the observable spectrum of the astrophysical source as if background were not present.
\AEacro\ implements an intuitive and straightforward background correction for standard quartiles, based on the observed cumulative distribution of the net spectrum, as shown in Figure~\ref{Emedian.fig}.
This method appears to be equivalent to that described by \citet[][Appendix~C]{Hong04}, which was developed independently.
\AEacro\ does not yet estimate individual confidence intervals for these background-corrected quartiles.

Our own \Chandra\ studies of young stars have chosen to use the median energy statistic, rather than multiple quartiles, for deriving intrinsic properties of absorbed thermal plasmas (\S\ref{photometry.sec}).
Many sources in these studies are so weak that dividing the few observed counts into multiple quartiles is not feasible.

%

\subsection{Photometry \label{photometry.sec}}

The common procedure for obtaining broad band source fluxes and luminosities (given astronomical distances) is to fit the spectrum derived above with astrophysical models convolved with the ARFs and RMFs representing the observatory response.  This path (which we follow in \S\ref{XSPEC.sec} for strong sources) is less reliable for faint sources with insufficient photons to constrain the nonlinear fitting process. 
We describe here a photometric procedure to estimate photon fluxes.

\AEacro\ begins by performing standard aperture photometry, computing ``net counts'' quantities for various energy bands, $S(E)$, by subtracting a scaled background from the counts observed near the source. 
\begin{eqnarray}
S(E) = C^s(E) - (A^s/A^b) C^b(E) \label{src_cts.eqn}
\end{eqnarray}
$C^s(E)$ is the number of counts observed in the source aperture in energy band $E$.
$C^b(E)$ is the number of counts observed in the background region in energy band $E$.
The source and background ``areas'', $A^s$ and $A^b$, are derived by integrating exposure maps rather than computing geometric areas (\S\ref{bkg_extract.sec}).
Note that these raw ``net counts'' photometric quantities are not adjusted to correct for finite apertures.
Instead, an energy-dependent aperture correction (\S\ref{aperturecorr.sec}) is applied to the calibration of the extraction (the ARF file); thus, any calibrated photometric quantities (such as the flux estimates below) will include the aperture correction.

Asymmetric confidence intervals for both $C^s$ and $C^b$  are estimated using the common analytical approximations to confidence intervals of the Poissonian distribution constructed by \citet[][equations 7 and 12]{Gehrels86}. 
These confidence intervals are propagated through equation~\ref{src_cts.eqn}, using the method described by \citet[][equation 1.31]{Lyons91}, to estimate an asymmetric confidence interval for $S(E)$. 
This technique is not ideal, and we anticipate adopting the Bayesian technique for estimating confidence intervals used by the \anchorfoot{http://cxc.harvard.edu/csc/why/ap_vals_errs.html}{\Chandra\ Source Catalog,} which is implemented in the \CIAO\ tool {\em aprates}.

These standard ``net counts'' quantities are used to construct two estimates of incident photon flux onto the \Chandra\ telescope, designated here $F^\star$ and $F^\#$, in units of photon~cm$^{-2}$~s$^{-1}$ (not the usual erg~cm$^{-2}$~s$^{-1}$).                              
For the first estimate, the net count rate is divided by the observatory response in each of many narrow channels, and these nearly-monochromatic photon flux densities are summed over a chosen energy band:
\begin{eqnarray}
F^\star_{phot}(E) &=& \frac{S(E)}{{\rm EXPOSURE} \times {\rm ARF}(E)} \\
F^\star_{phot}(E_{min}<E<E_{max}) &=& \sum_{E_{min}}^{E_{max}} F^\star_{phot}(E) \label{NetFlux1.eqn}
\end{eqnarray}
where $S(E)$ is the net observed counts (equation~\ref{src_cts.eqn}), ${\rm ARF}(E)$ is the observatory effective area (\S\ref{OGIP_files.sec}), ${\rm EXPOSURE}$ is the source exposure time, and $E_{min}$ and $E_{max}$ are the energy range provided by the user.\footnote{
A similar method is used in the \Chandra\ Source Catalog where it is called \anchorparen{http://asc.harvard.edu/csc/columns/fluxes.html}{``aperture source energy flux''}.
}
Commonly used energy bands are $E_{min}=0.5$~keV and $E_{max}=2$~keV for the \Chandra\ ``soft band'' and $E_{min}=2$~keV and $E_{max}=8$~keV for the \Chandra\ ``hard band.''   

For the second photon flux estimate, the net count rate in a user-supplied band is divided by the observatory response averaged over the band:
\begin{eqnarray}
{\rm ARF}(E_{min}<E<E_{max}) &=& \sum_{E_{min}}^{E_{max}} {\rm ARF}(E)
	\label{ARF_min_max.eqn} \\
F^\#_{phot}(E_{min}<E<E_{max}) &=& \frac{\sum_{E_{min}}^{E_{max}} S(E)}{{\rm EXPOSURE} \times {\rm ARF}(E_{min}<E<E_{max})} \label{NetFlux2.eqn}
\end{eqnarray}

The $F^\star$ estimator can suffer from large Poisson errors because events (either source or background) at energies where the ARF is very small have a large effect on the estimator, and is thus not recommended for weak sources.  For example, an event at 8~keV where the ARF value is tiny makes a much larger contribution to $F^\star$ than an event at 2~keV where the ARF value is large.  The $F^\#$ estimator suffers from a systematic bias with respect to the true incident photon flux because the effective area averaged over the energy band (equation~\ref{ARF_min_max.eqn}) is the correct calibration of the net counts photometry only for the non-physical case of a flat incident spectrum. 

When {\it a priori} information provides constraints on the shape of a source's intrinsic spectrum, these photon flux estimates $F^\star$ and $F^\#$ can be combined with the background-corrected median energy (\S\ref{Emed.sec}), $E_{median}$, and with the astronomical distance $d$ to estimate apparent source luminosities $L_x$ in a chosen broad energy band:
\begin{equation}
L_x \approx 4 \pi d^2 [F^\star {\rm ~or~} F^\#] E_{median}. \label{Lx.eqn}
\end{equation}
\citet{Getman10} has studied in detail the accuracy and precision of this estimate for both bright and faint sources, particularly when absorption from line-of-sight interstellar matter is present.  $E_{median}$ is an accurate, though nonlinear, predictor of interstellar column density $N_H$ when the intrinsic family of spectral models (e.g., power law, thermal plasma) for the source is known \citep[][Figure~4]{Feigelson05,Getman10}.  $E_{median}$ thus enters the calculation twice, once to scale the photon flux to the observed luminosity in equation (\ref{Lx.eqn}), and again to scale the observed luminosity to the intrinsic luminosity \citep[corrected for absorption;][Figure~5]{Getman10}.  

Simulations \citep{Getman10} indicate that these nonparametric estimates of observed luminosities and absorption
column densities are quite accurate, with only moderate biases and statistical errors. 
Systematic errors are larger for estimates of the absorption-corrected luminosities, and can be extremely large for heavily absorbed sources when the chosen spectral band includes the soft X-ray regime. 
In our work on faint sources, we choose to use the more statistically stable $F^\#$ photon flux estimator and prefer absorption-corrected luminosities in the hard band (2--8~keV) (which is less vulnerable to errors in absorption correction) over luminosities in the total band (0.5--8~keV).

\subsection{Spectral Fitting \label{XSPEC.sec}}

\begin{figure}[htb]
\centering
\includegraphics[width=1.0\textwidth]{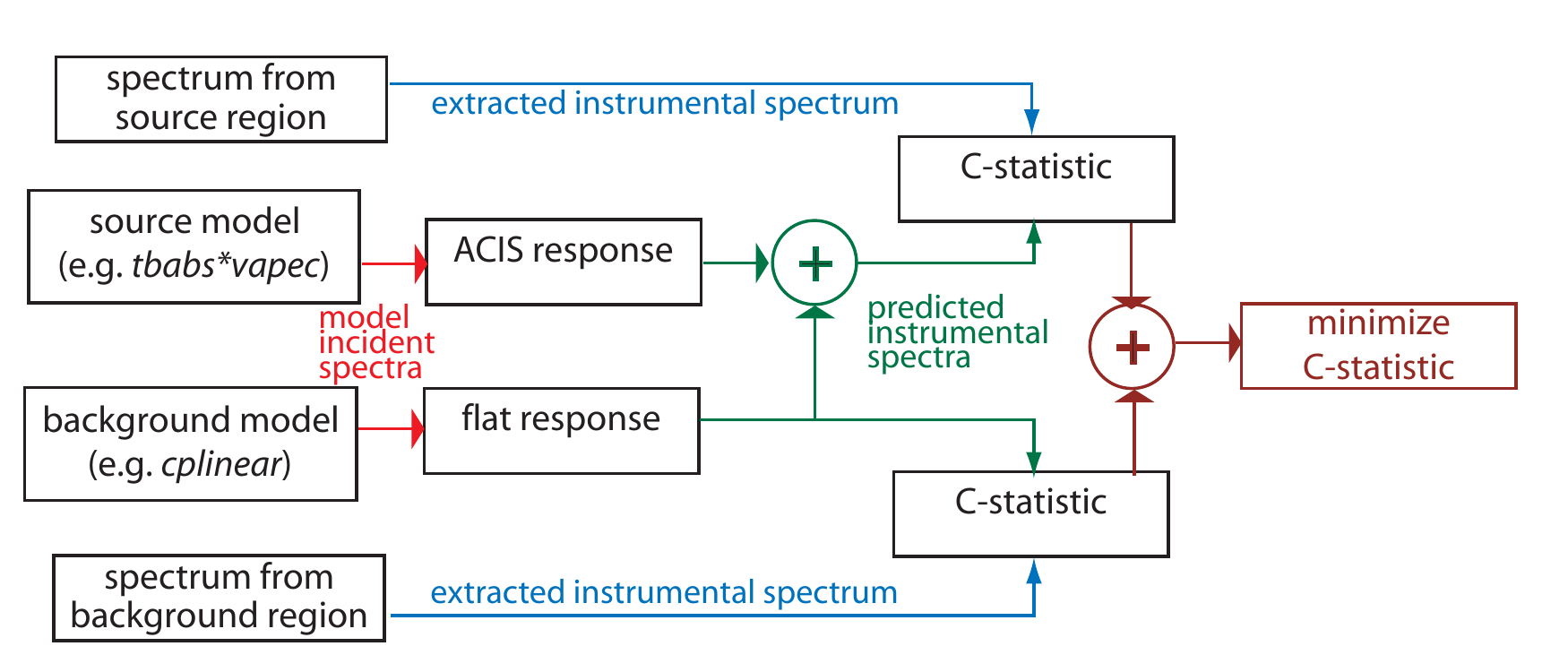} 
\caption{Data flow diagram of parametric fitting procedure using the C-statistic with the {\it cplinear} background spectral model (\S\ref{XSPEC.sec}).  The left plus sign represents the addition of two predicted instrumental spectra, and the right plus sign represents the addition of two C-statistic values corresponding to multiplication of the likelihood of the source and background data. 
\label{cplinear_dataflow.fig}}
\end{figure}

\AEacro\ relies on the widely-used \XSPEC\ package\footnote{
The \anchorparen{http://cxc.harvard.edu/sherpa/}{\Sherpa\ package} provides similar fitting capabilities, but the authors had more experience with \XSPEC\ at the time \AEacro\ development began.
}
\citep{Arnaud96} for spectral fitting.
\XSPEC\ provides a wide range of intrinsic spectral models (such as non-thermal power laws and thermal plasmas), absorption by intervening material, and convolution with the \Chandra/\ACIS\ instrumental response. 
Models derived from $\chi^2$ minimization become inaccurate for sources with few counts, and the problem is exacerbated when a significant fraction of the extracted events are background that is subtracted.  Spectral fitting of faint sources is thus usually pursued using the  C-statistic applied to unbinned data, an application of the Likelihood Ratio Test under the assumption that the data follow the Poisson distribution \citep{Cash79}.  
Background subtraction is not possible for likelihood-based statistical procedures; instead, a background model must be included in the spectral model that is fit to the data (source plus background) extracted from the source aperture, and that background model should be simultaneously fit to the data extracted from the background region.
This fitting procedure is depicted in Figure~\ref{cplinear_dataflow.fig} as a data flow diagram.

\XSPEC\ implements such a background model, which was derived by \citet{Wachter79}, for use with the C-statistic.\footnote{
See Appendix B of the \anchorparen{http://heasarc.gsfc.nasa.gov/docs/xanadu/xspec/manual/manual.html}{\XSPEC\ manual}.
}
This model consists of a free parameter representing the background flux in each of the hundreds of spectral channels used in an \ACIS\ spectrum.
Such a model raises two theoretical concerns.
First, in most cases where the C-statistic would be used, there are many more free model parameters than background events.
Second, such a model is completely unconstrained---extremely narrow and quite non-physical features (one channel wide) in the incident background spectrum are presumed to be possible.
Indeed, observers have found that this algorithm sometimes provides very poor ``best fits'' to the observed spectrum, with incorrect normalizations and non-physical spectral parameters.\footnote{
See \url{http://xspector.blogspot.com/2007/01/cstat-with-background-files-and-low.html} and \url{http://xspector.blogspot.com/2005/07/cstat.html}.
}
    
\begin{figure}[htb]
\centering
\includegraphics[angle=90,width=0.5\textwidth]{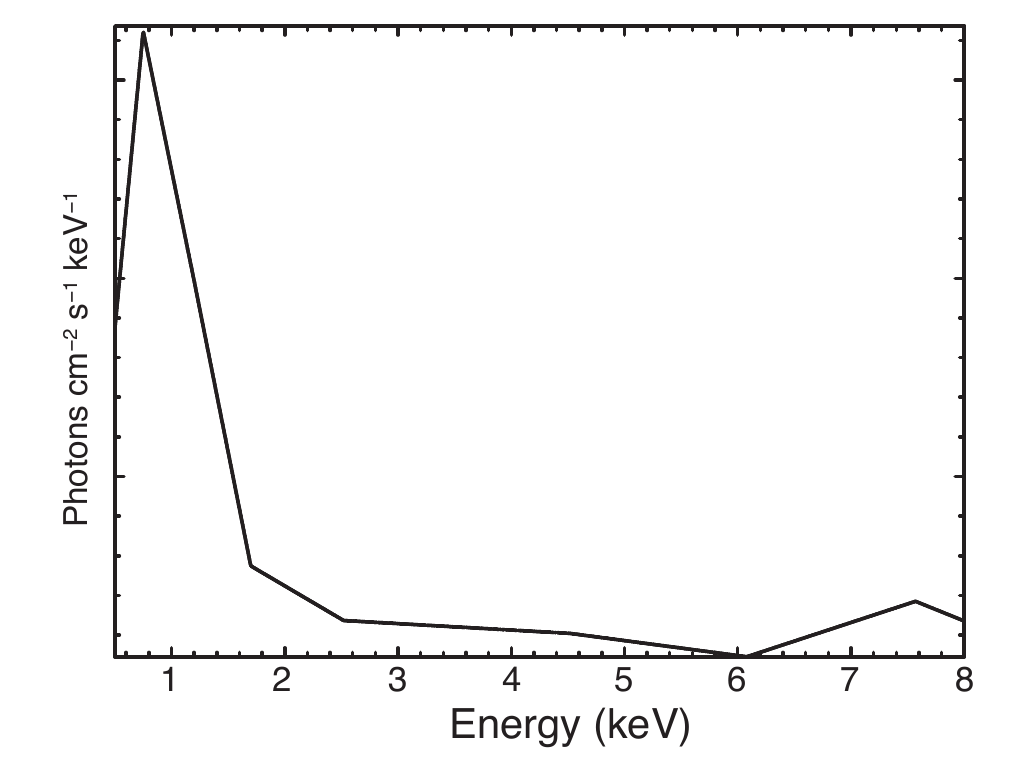} 
\caption{Example of a {\it cplinear} background spectral model (\S\ref{XSPEC.sec}) for a point source coincident with soft diffuse emission. 
The model shown has 8 vertices (7 linear segments).
The ordinate axis uses an arbitrary linear scale. 
\label{cplinear_example.fig}}
\end{figure}
\begin{figure}[htb]
\centering
\includegraphics[width=0.5\textwidth]{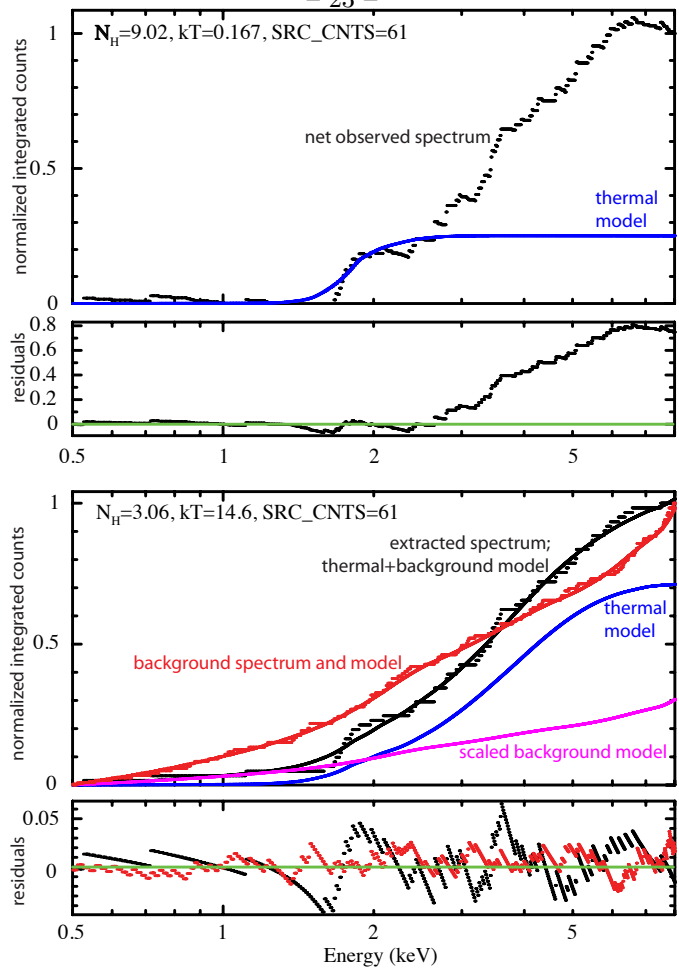} 
\caption{Example of spectral models for a faint stellar member of the M~17 star forming region derived with the C-statistic in \XSPEC\ using the \citet{Wachter79} (top panel) and {\it cplinear} (bottom panel) background models (see Figure~\ref{cplinear_dataflow.fig}).  
Residuals are shown at the bottom of each panel. 
The top panel shows the observed cumulative {\em net} spectrum (black stair-step curve) and the best-fit thermal plasma model (blue continuous curve) for the star; the Wachter background model is not shown. 
Note that the fit is incorrect above 1~keV.  
The bottom panel shows cumulative spectra from the source aperture and background region separately.
\XSPEC\ is configured so that the spectrum extracted from the background region (red stair-step curve) is compared to a {\em cplinear} model (red continuous curve), and the spectrum extracted from the source aperture (black stair-step curve) is compared to the sum (black continuous curve) of a scaled copy of the {\em cplinear} model (purple curve) and a thermal plasma model of the star (blue curve).
Best-fit parameters for the two models are then derived simultaneously.
\label{cplin_ex.fig}}
\end{figure}

We found that substituting a more constrained model for the \ACIS\ background stabilized the algorithm and lessened the problem of non-physical fits.  
For sources with relatively few (${\sim}100$) background counts, we use a continuous piecewise-linear function with up to 10 vertices, which we call the {\it cplinear} model;\footnote
{
See the discussion of the {\em cplinear} model in the \AEacro\ manual.
A comparison of source models using the \citet{Wachter79} and {\it cplinear} background models for 900 \ACIS\ sources is available in Appendix~B of that document.
} 
an example is shown in Figure~\ref{cplinear_example.fig}.
Figure~\ref{cplin_ex.fig} gives an example where the background model of \citet{Wachter79} gives a poor fit and the {\it cplinear} model gives a good fit.  

The highly constrained {\em cplinear} model is inadequate to model high quality background spectra which exhibit complex structure.
More physically-based background models would be useful to the \ACIS\ community; our goal here is to emphasize that severely under-constrained background models should be avoided, rather than to claim that {\em cplinear} is optimal.
The ultimate solution to modeling low-count X-ray spectra may lie in Bayesian approaches, such as the innovative technique of \citet{vandyk01}.
Those authors also suggest that background models with appropriate functional forms are useful, however the background model presented in that published work employed a free parameter for each channel in the spectrum \citep[eq. 22]{vandyk01}, similar to the Wachter model.

\begin{figure}[htb]
\centering
\includegraphics[width=1.0\textwidth]{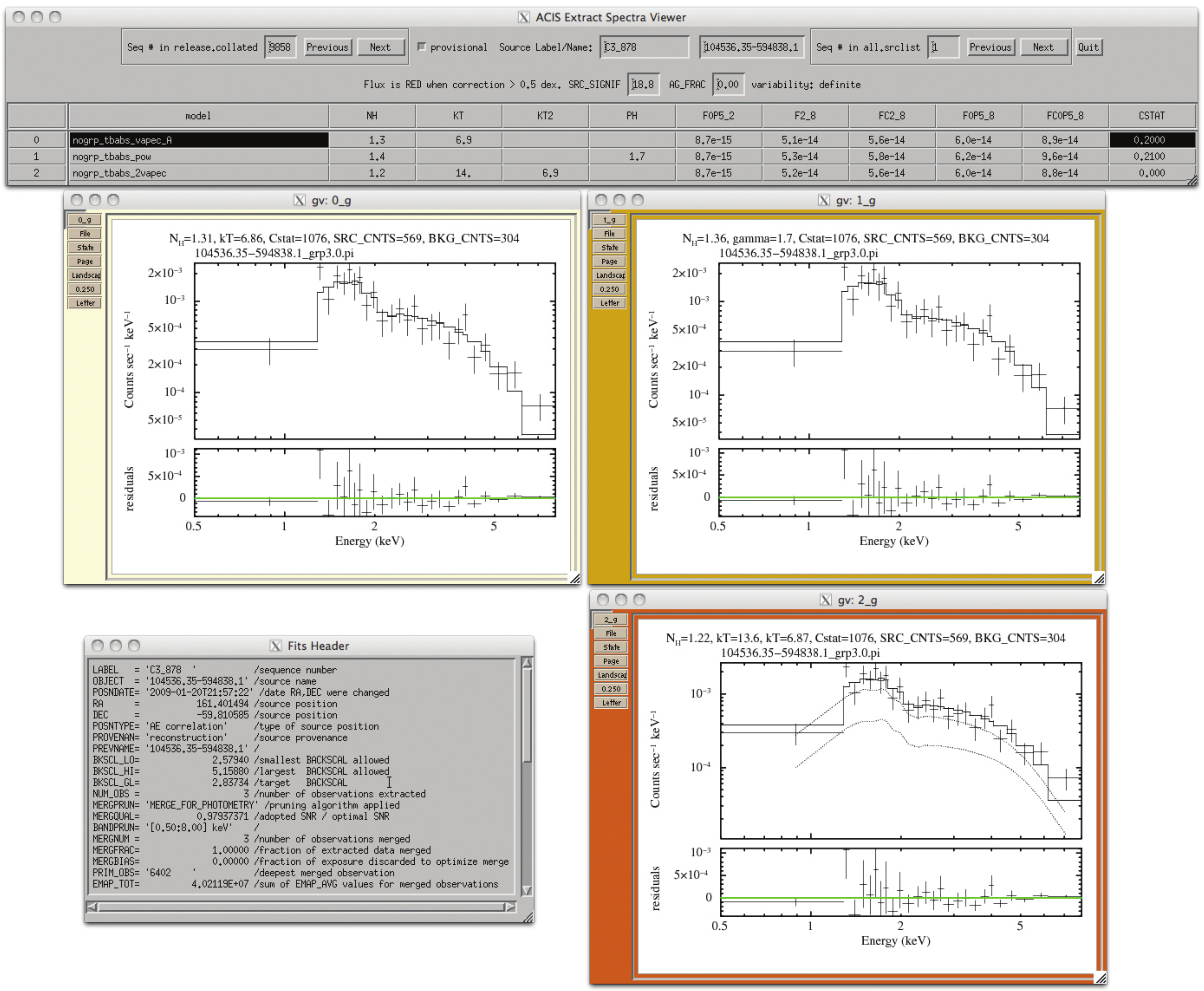} 
\caption{Screen shot of three spectral fits to an \AEacro\ source presented in the {\em Spectra Viewer} tool.
An IDL ``widget'' (top) provides navigation controls to select a source from the catalog, displays tabular information about all the spectral fits available for the selected source, and allows the observer to choose the preferred fit.
Standard \XSPEC\ plots from each spectral fit are shown in separate {\em ghostview} windows.
A FITS header containing various properties of the selected source is shown (lower left window).
\label{spectra_viewer.fig}}
\end{figure}

Another algorithmic innovation to spectral fitting provided by \AEacro\ is the improved method for grouping spectra (either for fitting with the $\chi^2$ statistic or for plotting ungrouped spectra fit with the C-statistic) described earlier (\S\ref{grouping.sec}). 
Finally, \AEacro\ also offers the observer several procedural conveniences.
\XSPEC\ scripts that drive fits to three common astrophysical models (absorbed one- and two-temperature thermal plasmas, absorbed power law), using either $\chi^2$ or the C-statistic with the {\it cplinear} background model, are provided.
\AEacro\ automates the execution of these fitting scripts; fitting multiple models to thousands of sources is feasible.
When multiple models have been fit to a source, the observer can visualize those models and select the most appropriate one using the graphical interface in the \AEacro\ tool {\em Spectra Viewer}; an example screen shot is shown in Figure~\ref{spectra_viewer.fig}.
Numerical output of the fitting process (in the form of a FITS table) includes best-fit spectral parameters with their confidence intervals and model fluxes integrated over various spectral bands. 
        

\subsection{Source Variability \label{variability.sec}}

\begin{figure}[htb]
\centering
\includegraphics[width=0.9\textwidth]{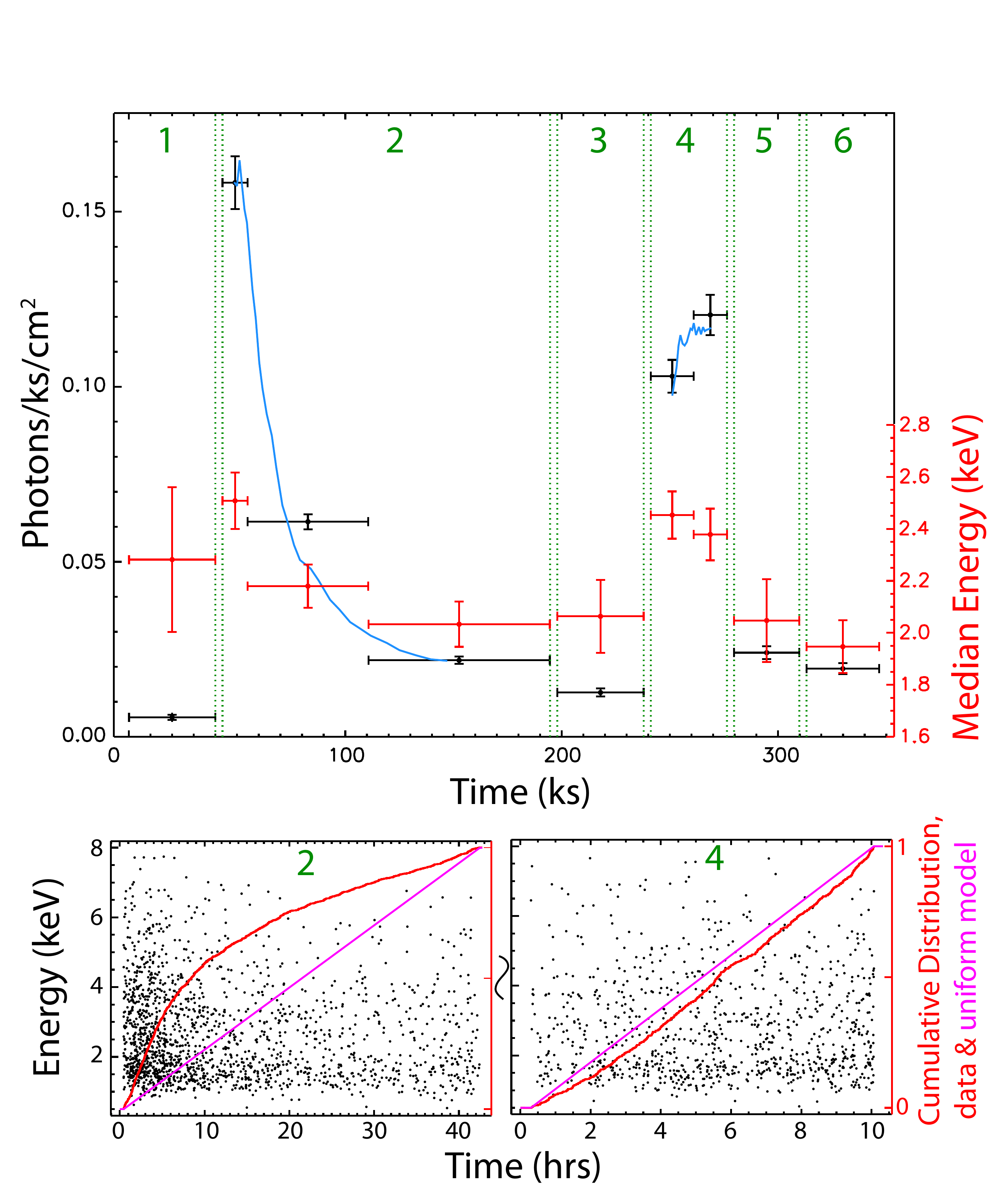} 
\caption{
Concatenated light curves for a source observed six times (six numbered epochs in upper panel) and photon arrival diagrams for two of those observations (lower panels).
The light curves show variations in photon flux (black histogram bins and blue adaptively smoothed curves) and median energy (red histogram bins); the spectrum is seen to harden when the source flares.
The photon arrival diagrams depict the arrival time and energy of each event (black dots) in a single observation, the cumulative distribution of event arrival times (red curve), and the cumulative distribution expected from a constant source (magenta curve).
\label{timing.fig}}
\end{figure}

For visualization of source variability, light curves for all observations of the source are depicted on a single plot, which is shown in calibrated units (photon~ks$^{-1}$~cm$^{-2}$) rather than observed units (count~ks$^{-1}$) to avoid spurious apparent variability arising from differences in observatory response or PSF fraction between observations.
Two light curves are constructed.
One is a histogram with variable-width time bins, constructed by the same algorithm used to group spectra (\S\ref{grouping.sec}).  
The second is adaptively smoothed using a box-kernel whose size is adjusted to encompass a specified number of events.
The adaptive smoothing process also produces a time series reporting the median energy of the observed events falling within each kernel, providing a visualization of spectral variation over time.
For each observation of a source, \AEacro\ also produces a ``photon arrival diagram''---a scatter plot of event arrival time and energy.
Figure~\ref{timing.fig} shows an example multi-observation light curve and a photon arrival diagram for a flaring source. 

For quantification of variability, the null hypothesis of a uniform source flux is tested using a one-sample Kolmogorov-Smirnov statistic.\footnote{
The \Chandra\ Source Catalog also provides \anchorparen{http://asc.harvard.edu/csc/columns/variability.html}{variability tests} based on the variance, Kuiper statistic, and \anchorparen{http://asc.harvard.edu/csc/why/gregory_loredo.html}{Bayesian method} of \citet{Gregory96}.
}
For multi-observation cases, variability is tested both within and between observations (accounting for variations in effective area among the observations).   

The current timing analysis has some deficiencies.  
No tests for autocorrelation, periodicity, or other temporal behaviors are made.  
The KS statistic is insensitive to variations near the beginning or end of the observation.
Variability in the background is not considered in the uniform flux model, and background is not accounted for in the light curves or median energy time series.
Although the \ACIS\ background is often quite stable within an observation, multiple observations of a source can easily suffer very different background count rates within the source apertures because the aperture sizes can be quite different.
Thus, for example, a spurious indication of variability can be produced for a weak source observed at two very different off-axis angles.
We plan to eliminate this problem by adding to the uniform light curve model independent background levels for each observation.

\subsection{Averaged or ``Stacked'' Properties of Source Samples \label{stack.sec}}

It is often desirable to ``stack'' a sample of similar sources to construct a summed spectrum with higher SNR than available from individual objects.  This can be performed either for sources detected in the \ACIS\ data or at locations of objects known only from independent observations.  Stacking many similar objects can thus give extremely long effective exposures and sensitivities, however stacking results can be dominated by the few brightest objects in the group.  This technique has been widely used in \Chandra\ deep extragalactic field studies \citep[e.g.,][]{Hornschemeier01, Brandt01, Lehmer07}. 

In \AEacro, stacking is achieved by relabeling extraction subdirectories of interesting sources to resemble extractions of distinct observations of a single source.  Then the extractions are merged (\S\ref{merging.sec}) and source properties are estimated normally (\S\ref{src_prop.sec}).

\subsection{Collation of Multi-source Tables \label{multisource.sec}}

The properties of individual sources obtained by \AEacro\ can be collated into a summary FITS table with $>100$ columns and a row for each source.  The scientist can then study relationships among quantities using, for example, the {\it fv} or {\it chips} FITS viewers, \IDL\ two- and three-dimensional plotting, and \Racro\footnote{
\Racro\ is a large, \anchorparen{http://www.r-project.org}{public-domain} statistical analysis software package.} 
multivariate statistical analysis.  \AEacro\ also prepares standard quantities for publication as {\it LaTeX} tables, illustrated by Tables~\ref{src.tab} and \ref{spec.tab}.   This feature is particularly valuable for fields with hundreds or thousands of sources.

\section{Matching Point Source Catalogs \label{matching.sec}}

The task of ``matching'' point source catalogs---identifying entries in two or more catalogs that are believed to represent the same physical source---is fundamental to astronomical research, and is growing increasingly more common as publicly available and high quality catalogs covering many wavebands proliferate.
As yet, no standard algorithm for this task has been adopted by the community, and thus no single software tool for matching is in widespread use.

We currently use a matching tool written by one of the authors, the \IDL\ program {\it match\_xy} in \TARA\ \citep{Broos07}.
The basic criterion used to evaluate whether a proposed pair of sources from two catalogs could be detections of the same physical object is very simple.
Given a significance threshold chosen by the observer, Gaussian positional uncertainties specified individually for the two sources are used in a classical test of the null hypothesis that they are observations of the same object.
This matching criterion does not represent an innovation; the astronomy literature contains several more sophisticated catalog matching algorithms, such as ones that provide individual probabilities that each asserted match is merely a chance coincidence \citep{Downes86, Sutherland92, Rutledge00, Bauer00}, ones that consider non-positional source properties (such as fluxes) \citep{Brand06, Budavari08}, and ones that operate on more than two catalogs simultaneously (e.g., the \anchorfoot{http://openskyquery.net/Sky/SkySite/help/algo.aspx}{{\it XMatch} algorithm} in the {\it SkyQuery} service designed for the Virtual Observatory).

However, we do recommend four features of our matching tool for use in any catalog matching procedure applied to \Chandra\ sources.
First, individual estimates of positional uncertainty for each source, rather than a canonical value for the whole catalog, are employed because they often span a wide range due to \Chandra's variable PSF; the uncertainties estimated by \AEacro\ (\S\ref{src_positions.sec}) for a recent project range 
from $<0.1$\arcsec\ (not including systematic errors) to $>2$\arcsec.
Second, our algorithm enforces a one-to-one relationship among the pairs of sources declared to be ``successful'' matches, i.e., no source participates in multiple successful matches.
The algorithm reports a ``failed'' match when a pair of sources pass the matching criterion, but one of them is already participating in another (more reliable) match.
Third, we have found that visual depictions of the matching results, such as the \DSnine\ region files produced by {\it match\_xy}, are invaluable aids to reviewing the general performance of the algorithm and to understanding specific sources.
We have used these region files in a prototype {\em Source Viewer} tool that facilitates visual examination of the vicinity of each source in two or more images (e.g., X-ray and near-infrared), and records the observer's subjective comments on individual sources.
Fourth, all matching procedures should measure and correct astrometric offsets between catalogs at the outset; any large astrometric uncertainty should be included in the individual estimates of positional uncertainty.

\section{ANALYSIS OF DIFFUSE X-RAY STRUCTURES \label{diffuse.sec}}

Extended emission from hot diffuse plasma is commonly present in X-ray studies.  This occurs in galaxy groups and clusters, elliptical and spiral galaxies, Galactic star formation and starburst regions, supernova remnants, planetary nebulae, and wind-blown bubbles.  
Since point sources are often superposed onto the diffuse emission, the first step in our diffuse analysis workflow (Figure~\ref{workflow.fig}) is to identify and mask (remove) as many point sources as possible, using the procedures described in \S\ref{source_det.sec}--\ref{bkg_extract.sec}.
Observers must be aware that fainter undetected point sources may be present in some fields; interpretation of diffuse emission should consider this corruption by undetected point sources.     
Diffuse analysis proceeds with smoothing the resulting source-free image, defining regions that contain diffuse structures of interest, and extracting the X-ray data from those regions.

\subsection{Smoothing of Source-Free Images \label{smooth.sec}}

As described in \S\ref{bkg_extract.sec} and Figure~\ref{emap.fig}, \AEacro\ provides two methods for constructing mask regions that remove point source photons that would contaminate an analysis of diffuse emission.  
The remaining event distribution, consisting of instrumental background and possible astrophysical diffuse emission, is often best studied in smoothed images.
Although the \CIAO\ task {\it csmooth} \citep{Ebeling06} is effective and popular for smoothing unmasked data that contain both point sources and diffuse emission, it is poorly suited to working with images that contain unobserved regions, either masked point sources or regions beyond the edges of the detector, because the tool does not allow a field mask to be supplied.
The \CIAO\ \anchorfoot{http://asc.harvard.edu/ciao/threads/diffuse_emission/}{thread for diffuse emission} with embedded point sources replaces pixel values in source regions of an image with values interpolated from surrounding background regions using random Poisson numbers or local polynomial regression.

We prefer to apply an adaptive kernel smoothing tool in \anchor{http://www.astro.psu.edu/xray/acis/acis_analysis.html}{\TARA} that handles excised pixels and field edges, which is similar to the {\em asmooth} tool in the  \anchorfoot{http://xmm.vilspa.esa.es/external/xmm_sw_cal/sas_frame.shtml}{{\em XMM-SAS}} package and the new \anchorfoot{http://cxc.harvard.edu/ciao/ahelp/dmimgadapt.html}{{\em dmimgadapt}} tool in \CIAO\ 4.2.
Three smoothing kernels are available: a  top-hat (or Heaviside function), a symmetric two-dimensional Gaussian, and a two-dimensional Epanechnikov kernel \citep{Silverman92}.
The observer supplies a threshold SNR value; at each pixel position, the algorithm finds the smallest kernel that achieves this SNR target and estimates an incident photon flux (photon~s$^{-1}$) by dividing the number of counts under the kernel by the integrated exposure map under the kernel.
Maps of photon flux, flux error, and kernel size are produced.
The observer can choose to retain or smooth over the holes in the data introduced when point sources are masked.
The method is described in detail, with sample images, in \citet[][Appendix C]{Townsley03}.

\subsection{Extracting Diffuse Sources \label{diffuse_extract.sec}}

As the \AEacro\ package does not attempt to delineate the morphology of diffuse structures, the observer must define each diffuse source of interest in the form of a \DSnine\ region file that is accepted by \CIAO.
We have found the \anchorfoot{http://www.phy.ohiou.edu/~diehl/WVT/}{WVT Binning algorithm} \citep{Diehl06}, which is a generalization of the Voronoi binning algorithm described by \citet{Cappellari03}, to be a very effective method for tessellating a field with compact regions of similar SNR.\footnote
{
We can provide the reader with a small tool that converts the output of the WVT Binning package into a set of \DSnine\ region files.
}
Figure~\ref{diffuse.fig} shows an example if its use.

The procedure for extracting each diffuse source with \AEacro\ is analogous to the point source procedure.
For each observation, events within the source region are extracted, calibration products are constructed, and background spectra are extracted.
All the extractions are merged, photometric quantities are estimated, and spectra are fit via automated scripts.
However, more complex procedures are required to calibrate diffuse extractions and to account for background, as described in the next two sections.

\subsection{Calibration of Diffuse Extractions \label{exposure.sec}}

The \anchorfoot{http://cxc.harvard.edu/ciao/dictionary/arf.html}{ARF data product} constructed for point sources by the tool {\em mkarf} can be thought of more abstractly as an observatory response function, ${\rm ARF}(E,\mathbf{p})$, computed at a position on the sky, $\mathbf{p}$, for a monochromatic energy $E$, with units of cm$^2$~count~photon$^{-1}$.\footnote
{
An ARF is a product of a mirror effective area with units of cm$^2$ and a detector quantum efficiency with units of count~photon$^{-1}$. 
}
For \Chandra\ data, this function represents both variations in the observatory response across the focal plane and the exposure time variations across the sky caused by dithering over bad detector pixels and detector edges.\footnote
{
The \Chandra\ convention is that the exposure time (FITS keyword ``EXPOSURE'') recorded for a point source is always the nominal exposure time for the observation.  If necessary, a source's ARF is reduced to account for the amount of time the source was not observed due to dither motion.
}
Essentially, ${\rm ARF}(E,\mathbf{p})$ describes the ``depth'' of the observation, since effective area and observing time are degenerate terms in a flux calculation.
For example, ${\rm ARF}(E,\mathbf{p})$ trends downward with off-axis angle, due to vignetting of the mirrors.
${\rm ARF}(E,\mathbf{p})$ is reduced at any sky position that dithered across bad pixels or dithered off of \ACIS.
Conceptually, ${\rm ARF}(E,\mathbf{p})$ is zero in regions on the sky that have been masked to remove point sources (\S\ref{bkg_extract.sec}).

\begin{figure}[htb]
\centering
\includegraphics[width=0.5\textwidth]{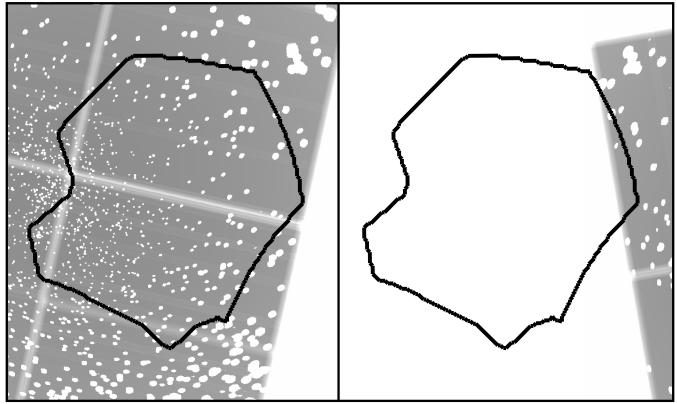} 
\caption{
A diffuse extraction region (black polygon) that is fully observed by one ACIS-I observation (masked exposure map, left panel) and partially observed by another observation (masked exposure map, right panel).
\label{diffuse_region.fig}}
\end{figure}

The ${\rm ARF}(E,\mathbf{p})$ function for any particular observation will vary within a diffuse extraction region, and in a multi-observation study may be zero over much of that region.
Figure~\ref{diffuse_region.fig} shows an example diffuse region that is fully contained in one ACIS-I observation (left panel) but barely intersects another ACIS-I observation (right panel).
Just as a diffuse extraction can be viewed as an {\em integration} of the observed counts over the extraction region, the proper calibration of that extraction can be viewed as an integration of ${\rm ARF}(E,\mathbf{p})$ over the region.
\begin{equation} 
{\rm ARF}_R(E) = \int_R {\rm ARF}(E,\mathbf{p})\,d\mathbf{p}  \label{ARF_R.eqn} 
\end{equation} 

Note that this integration over a region on the sky gives ${\rm ARF}_R(E)$ units that include a geometric area term, such as cm$^2$~count~photon$^{-1}$~arcsec$^2$.
As ${\rm ARF}_R(E)$ is carried forward into photometry calculations and spectral fitting, all ``flux'' quantities derived should be interpreted as surface flux quantities, with arcsec$^{-2}$ appended to the units.
Flux integrated over the diffuse region is then estimated by multiplying the surface flux by the region's {\em total} geometric area on the sky; the area lost to point source masks, bad pixels, and detector edges is already accounted for in ${\rm ARF}_R(E)$.

Recasting the diffuse ARF of each extraction to include the notion of ``area on the sky'' is particularly convenient when multiple extractions are to be combined (i.e., merged by \AEacro, just as is done with point sources).
Just as each extraction of a point source may have a different PSF fraction (\S\ref{aperturecorr.sec}), each extraction of a diffuse source may have a different ``area on the sky'' (due to masking and detector boundaries).
In both situations, corrections to the calibration are most intuitively applied to each extraction prior to merging.

${\rm ARF}_R(E)$ cannot be directly calculated with existing tools.
The \CIAO\ point source tool {\em mkarf} implements most of the behavior of the integrand ${\rm ARF}(E,\mathbf{p})$ described above, but has no mechanism to set ${\rm ARF}_R(E,\mathbf{p})$ to zero at positions where the event data have been masked.
The \CIAO\ tool {\em mkwarf} (``make weighted ARF'') implements an effective area averaging calculation that is related to Equation~\ref{ARF_R.eqn}, however for each observation {\em mkwarf} averages over only the {\em detector} area that intersects the diffuse region, not over the entire diffuse region, which may include areas with zero response due to masking or detector boundaries.
The difference between the diffuse ARF we have described here, ${\rm ARF}_R(E)$, and the data product returned by {\em mkwarf} can be summarized with the aid of Figure~\ref{diffuse_region.fig}.
For both extractions shown there, the {\em mkwarf} data products will have similar normalizations, because responses are averaged over the detector areas intersecting the extraction region.
In contrast, ${\rm ARF}_R(E)$ will be much larger for the left-hand extraction than for the right-hand extraction, because responses are averaged over the entire region.

Although the full ${\rm ARF}_R(E)$ function is not available as a standard data product, the value of ${\rm ARF}_R(E)$ at a single energy $E_0$ is easily calculated by integrating over the extraction region a standard \CIAO\ exposure map built for a monochromatic energy $E_0$ and masked in the same way as the event data. 
In the context of an \AEacro\ extraction, the exposure map is ``un-normalized'' with units of s~cm$^2$~count~photon$^{-1}$, built by supplying the ``normalize=no'' option to the \CIAO\ tool {\em mkexpmap}.
This exposure map is related to the integrand ${\rm ARF}(E,\mathbf{p})$ as
\begin{equation} 
{\rm emap}_{E_0}(\mathbf{p}) = {\rm EXPOSURE} \times {\rm ARF}(E_0,\mathbf{p}),
\end{equation}
where ${\rm EXPOSURE}$ is the ``exposure time'' of the observation.
Thus, the diffuse ARF we seek is easily calculated at one energy as
\begin{equation} 
{\rm ARF}_R(E_0) = \frac{\int_R {\rm emap}_{E_0}(\mathbf{p})\,d\mathbf{p}}{{\rm EXPOSURE}}  \label{emap_integral.eqn} 
\end{equation} 

With the normalization of ${\rm ARF}_R(E)$ established at energy $E_0$, \AEacro\ then relies on the {\em mkwarf} result to establish the shape of ${\rm ARF}_R(E)$ as a function of energy.
\AEacro's complete calculation of ${\rm ARF}_R(E)$ is thus
\begin{equation}
{\rm ARF}_R(E) = \frac{{\rm ARF}_{mkwarf}(E)}{{\rm ARF}_{mkwarf}(E_0)} \times \frac{\int_R {\rm emap}_{E_0}(\mathbf{p})\,d\mathbf{p}}{{\rm EXPOSURE}}  \label{diffuse_ARF.eqn} 
\end{equation} 
where ${\rm ARF}_{mkwarf}$ is constructed by {\em mkwarf}.

To summarize this section, we calibrate each diffuse extraction by constructing ${\rm ARF}_R(E)$, which has units of cm$^2$~count~photon$^{-1}$~arcsec$^2$ and accounts for the area on the sky lost to point source masks, bad pixels, and detector edges.
``Flux'' quantities subsequently derived should be interpreted as surface flux quantities, with arcsec$^{-2}$ appended to the units.
Flux integrated over the diffuse region is estimated by multiplying the surface flux by the region's {\em total} geometric area on the sky.

\subsection{Background in Diffuse Extractions}

Diffuse sources are often heavily contaminated by background---both instrumental background and emission from foreground and background astrophysical sources that are not of interest.
Several strategies to account for background in diffuse sources are described in the \AEacro\ manual.
We choose to subtract from each diffuse extraction a standard instrumental background spectrum obtained by scaling the so-called \anchorfoot{http://cxc.harvard.edu/ciao/threads/acisbackground/index.py.html}{``\ACIS\ stowed event data''} provided in the \Chandra\ Calibration Database \citep{Hickox06}.
After merging the multiple observations of a diffuse region (with the same methods that are used for point sources in \S\ref{merging.sec}, with no optimization options enabled), standard ``source'' and ``background'' spectra with calibration data products are available.
Astrophysical background contaminating a diffuse region can be handled in either of two ways.
First, if a nearby ``sky'' region thought to be mostly free from the emission under study is available, then its observations and ``stowed backgrounds'' are extracted, and two net spectra from the diffuse region and the sky region are simultaneously fit using a shared model for the astrophysical background in both regions plus a source model for the emission of interest in the diffuse region.
Alternatively, if no suitable ``sky'' region is available, then the astrophysical background must be directly modeled \citep{Snowden08}.
If appropriate fitting scripts are constructed, \AEacro\ can automate this spectral fitting in the same way that point source spectral fitting is automated.

\subsection{Visualizing Model Parameters for Spectra of Diffuse Structures \label{diffuse_maps.sec}}

\begin{figure}[htb]
\centering
\includegraphics[width=0.95\textwidth]{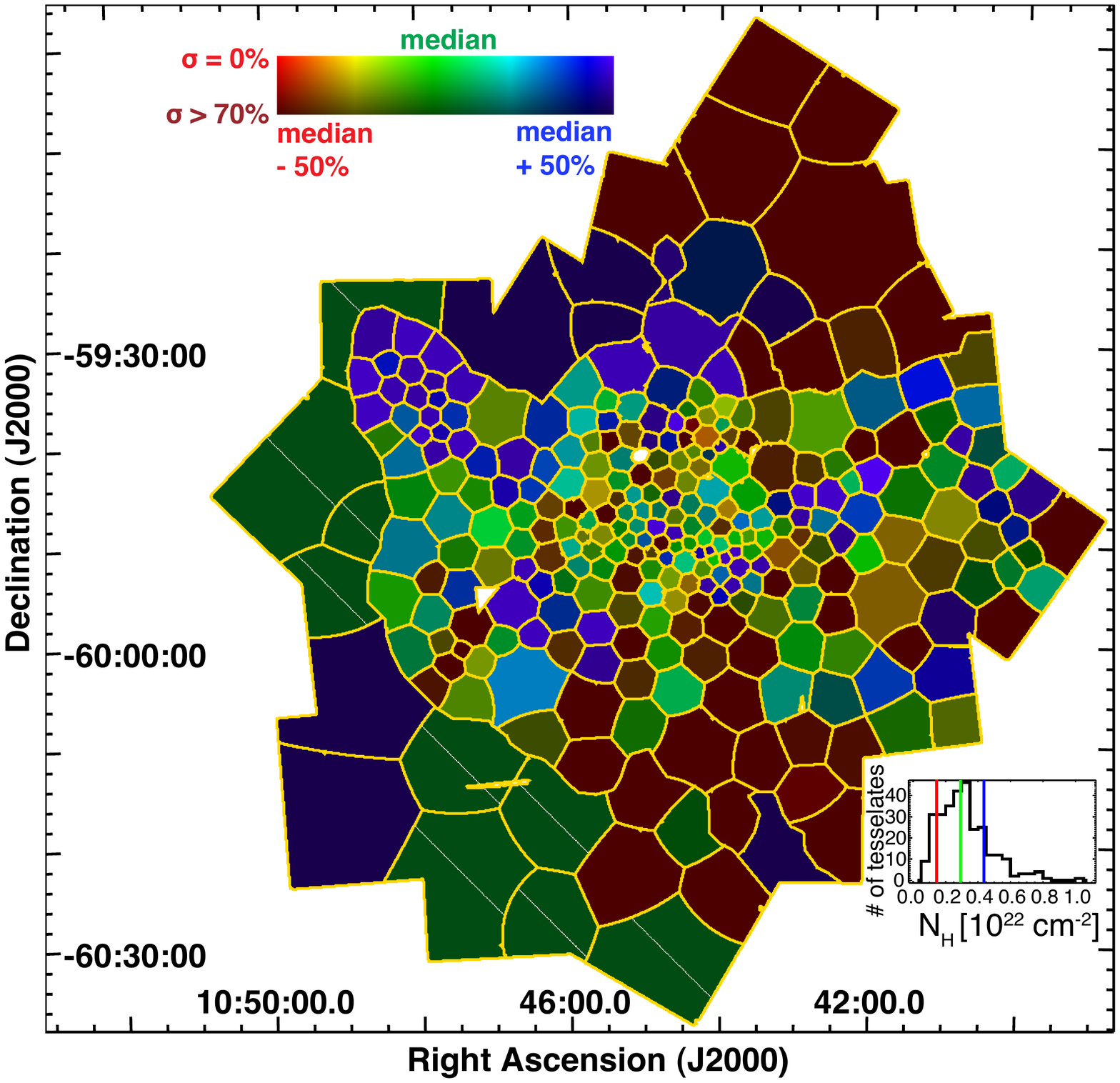} 
\caption{
\AEacro\ analysis of the Chandra Carina Complex Project \citep{Townsley10} study of the Carina Nebula (see Figure~\ref{emap.fig}), a region with strong diffuse emission with complex morphology.
Diffuse regions (yellow) are defined by tesselating an image of diffuse emission using the WVT Binning algorithm \citep{Diehl06}.
In this example map the color inside each region represents the absorption parameter ($N_H$) from the spectral model derived from the corresponding extracted spectrum.
As shown in the legend, the hue (red ... blue) of the color encodes the $N_H$ {\em value} (relative to the median):
red hues represent values 50\% below the median; green hues represent the median value, 0.29; blue hues represent values 50\% above the median.
The brightness of the color encodes the {\em uncertainty} of that value.
For example, a highly certain low $N_H$ value would be bright red; a highly uncertain low value would be maroon.  
Regions where the parameter was frozen in the fit are marked with white diagonal lines.
The inset plot shows the distribution of mapped $N_H$ values ( in units of $10^{22}$~cm$^{-2}$) with the values corresponding to the red, green, and blue hues marked by vertical lines. 
\label{diffuse.fig}}
\end{figure}

When more than a few diffuse regions are defined, many observers have found that spectral fitting results can be best understood by creating maps that show various results, e.g., model parameters and fluxes.
Standard gray-scale or false-color maps effectively depict to the human eye spatial variations in a parameter.
We are currently experimenting with a complementary map-coloring technique that seeks to depict both a parameter's value and its uncertainty, using hue to encode the parameter value and brightness to encode its uncertainty, as shown in Figure~\ref{diffuse.fig}.  

\section{VISUALIZATION  \label{visualiz.sec}}

Visualization of data products throughout the analysis workflow shown in Figure~\ref{workflow.fig} can reveal various problems that commonly arise, including artifacts in the data, mistakes in execution of the workflow such as skipping a step or failing to recognize a failure, bugs in software tools or unexpected changes in the algorithms implemented by tools, astrometric offsets in observations, or uncommon features in the data such as CCD readout streaks, severely piled-up sources, and sources lying on the edge of or just outside the field of view.
We display the events removed from the observation at each cleaning step of the L1-to-L2 processing (\S\ref{L1_L2.sec}) to look for unexpected patterns.
We plot candidate point sources arising from the detection process (\S\ref{cand_src.sec}) on top of the event data to look for duplicates and spurious detections.
As the candidates are extracted by \AEacro, each observation's source apertures are examined in \DSnine\ to verify that they are not overlapping (\S\ref{extract_reg.sec}).  
If \AEacro\ masks the data to remove point sources (\S\ref{bkg_extract.sec}), then the resulting background data set is scanned to determine if additional masking is required (e.g., around very bright sources).
Candidate sources that are found to be not significant (likely background fluctuations) are reviewed before they are pruned from the catalog (\S\ref{iterative_detection.sec}).
All \AEacro\ source position estimates (\S\ref{src_positions.sec}) that differ significantly from the original position assigned by the detection process are visually confirmed before the revised position is adopted.

Many visual reviews consist of displaying a project-level event list (constructed by merging all the observations) in \DSnine, overlaid with color-coded \DSnine\ regions depicting sources of interest.
\AEacro\ provides a complementary tool (the ``SHOW'' option) that examines a single source in detail, as shown in Figure~\ref{PSF_polygons.fig}.
The neighborhood around the source in each observation is shown in a separate \DSnine\ frame, overlaid with the extraction aperture; another frame shows the merged neighborhood.
If a reconstructed image of the merged neighborhood is available, then it is shown in an additional frame.
Basic information about each extraction is presented in tabular form.

After the catalog is extracted and source properties are collated, we plot the distributions of key source properties, looking for outliers that signify mistakes (or discoveries).
Light curves are examined for sources exhibiting strong variability (\S\ref{variability.sec}).
Results from spectral fitting are reviewed using a custom tool designed for that purpose (\S\ref{XSPEC.sec}).

\section{SUMMARY \label{discussion.sec}}

Many \Chandra-\ACIS\ imaging studies face significant data analysis challenges arising from large numbers of weak and sometimes crowded point sources embedded in scientifically relevant diffuse emission, observed with multiple misaligned pointings. 
We have discussed here a variety of innovations to standard \ACIS\ analysis methods that address these challenges; the most important of these are summarized below.
\begin{enumerate}
\item Currently, a single set of data cleaning procedures is not adequate if the observer plans to study both very weak (${\leq}10$ counts) point sources (or diffuse emission) and bright point sources, because several cleaning algorithms  remove legitimate X-ray events from bright sources.
Thus, we find that distinct data cleaning procedures are required for different types of analyses (\S\ref{L1_L2.sec}). 

\item When point sources are to be extracted, we recommend evaluating the existence of sources using those extraction results rather than relying on typical source detection tools for that judgment (\S\ref{iterative_detection.sec}). 

\item In crowded fields we recommend searching for candidate sources in reconstructed images (\S\ref{cand_src.sec}), defining extraction apertures that do not overlap (\S\ref{extract_reg.sec}), and defining background regions that seek to model the background contributed by the wings of nearby sources (\S\ref{bkg_extract.sec}).

\item We recommend correcting all point source extractions for the energy-dependent fraction of light that lies outside the source aperture (\S\ref{aperturecorr.sec}).

\item When a source is observed multiple times, we recommend estimating source validity, position, and photometry using three independent combinations of the extractions, each allowed to discard observations to optimize the accuracy of the corresponding quantity (\S\ref{composite_discard.sec}). 

\item We offer an algorithm for grouping spectra (\S\ref{grouping.sec}) that lessens biases inherent in standard algorithms.

\item We raise concerns about employing under-constrained background models in the context of spectral fitting and offer an alternative (\S\ref{XSPEC.sec}).  

\item When diffuse X-rays are extracted from a region that contains unobserved areas (e.g., due to detector edges or point source masks) we describe a necessary correction to the calibration provided by the tool {\em mkwarf} (\S\ref{exposure.sec}).

\end{enumerate}

Most of the data analysis methods we have discussed are implemented in the \anchor{http://www.astro.psu.edu/xray/acis/acis_analysis.html}{{\em ACIS Extract} (AE) software package,} which has been freely available to the community since its development began in 2002.\footnote
{
Development of AE is on-going; observers can be notified of new releases by joining the \anchorparen{http://lists.psu.edu/cgi-bin/wa?A0=L-ASTRO-ACIS-EXTRACT}{AE mailing list}.
}
\AEacro\ is written in the \IDL\ language, and makes extensive use of tools in \CIAO\ and in several other public packages.

Although much of the analysis we perform on our \ACIS\ observations has been automated, we believe that the obserer should retain many vital roles in the process.
The human eye is often able to spot omissions and spurious entries in the set of candidate sources derived from detection procedures (\S \ref{cand_src.sec}).
We rely on the observer to judge what effect CCD readout streaks may have on the data analysis and to take appropriate mitigating actions (\S \ref{bkg_extract.sec}).
Scientific judgement is required to select appropriate levels of smoothing for diffuse sources (\S \ref{smooth.sec}) and to define appropriate regions to extract (\S \ref{diffuse_extract.sec}).
We encourage the observer to visually review (\S \ref{visualiz.sec}) data cleaning steps, extraction apertures, catalog pruning and source repositioning proposed by algorithms, spectral fitting results, and multi-wavelength associations asserted by our matching algorithm.

\acknowledgments{The authors greatly appreciate discussions with Keith Arnaud regarding background modeling when the C-statistic is used in \XSPEC, and several good suggestions for expanding \AEacro\'s capabilities from Mike Muno.   
We would have enjoyed considerably less success in our studies of diffuse sources had Steven Diehl and Thomas Statler chosen to not release their WVT Binning software to the community; we thank them for this valuable service.
We appreciate the time and useful suggestions contributed by our anonymous referee.
We would have been lost without the invaluable tools of NASA's Astrophysics Data System, and without \CIAO, \XSPEC, and \DSnine.

This work is supported by the \ACIS\ Instrument Team contract SV4-74018 (PI: G. Garmire), issued by the Chandra X-ray Center, which is operated by the Smithsonian Astrophysical Observatory for and on behalf of NASA under contract NAS8-03060.  Additional support comes from Chandra General Observer grants 
GO8-9131X,
GO8-9006X,
GO6-7006X,
GO5-6143X, 
GO4-5006X, 
several previous Chandra General Observer grants, 
the NASA grant NNX06AE56G, 
and the NASA/ADP grant NNX09AC74G.}

\facility{CXO (ACIS)}

\appendix

\section{RUNNING TWO AFTERGLOW ALGORITHMS \label{spare_bit.sec}}
We concur with the \anchorfoot{http://cxc.harvard.edu/ciao/threads/acishotpixels/index.html\#caveats}{CXC's recommendation} to use both afterglow tools ({\em acis\_detect\_afterglow} and {\em acis\_run\_hotpix}) for identifying afterglow events (\S\ref{L1_L2.sec}) when seeking to detect weak sources.
Since both algorithms use bit 16 in the \anchorfoot{http://space.mit.edu/CXC/docs/docs.html\#evtbits}{event STATUS word}
to represent afterglows, the operational details of applying both algorithms can be confusing.
We show below one method for doing this.
First, the afterglow flag returned by the gentle algorithm in the tool {\em acis\_run\_hotpix} is moved from bit 16 to the unused bit 31 ({\em dmtcalc} call below).
Second, the aggressive algorithm in the tool {\em acis\_detect\_afterglow} is applied, storing its results in bits 16--19.
(Note that in the {\em dmtcalc} syntax below bits are numbered 0 to 31, counting from right to left.)  

\begin{verbatim}
  dmtcalc infile=acis.evt1 outfile=temp.evt \
          expression="status=status,status=X0F,if(status==X15T)then(status=X0T)"

  acis_detect_afterglow infile=temp.evt outfile=acis.evt1 \
                        pha_rules=NONE fltgrade_rules=NONE 
\end{verbatim}

Gentle cleaning that does not damage bright point sources (\S\ref{L1_L2.sec}) can be performed on the resulting STATUS word by ignoring the bits assigned by the {\em destreak} tool (bit 15), the {\em acis\_detect\_afterglow} tool (bits 16--19), and the Very Faint Mode grading algorithm (bit 23): 

\begin{verbatim}
  dmcopy "acis.evt1[status=00000000x000xxxxx000000000000000]" gently_cleaned.evt
\end{verbatim}

Aggressive cleaning that is suitable for source detection and for analysis of diffuse emission (\S\ref{L1_L2.sec}) can be performed by requiring all STATUS bits to be zero:

\begin{verbatim}
  dmcopy "acis.evt1[status=0]" aggressively_cleaned.evt
\end{verbatim}

\section{Source Significance \label{signif.app}}

\AEacro\ provides two measures that quantify the validity of a source.
First, a traditional ``signal-to-noise ratio'' is defined to be the ratio of the net source counts observed to the estimated uncertainty on that quantity.

Second, the null hypothesis that a source does not exist in the source aperture is tested by the method described by \citet[][Appendix~A2]{Weisskopf07}, which is derived under the more physical assumption that X-ray extractions follow Poisson distributions.
Assuming the null hypothesis---all the events in the source aperture are background---they show that the joint probability of finding at least the observed number of events in the source aperture and finding the observed number of events in the background region can be found by integrating a binomial distribution.  
This calculation can be performed with the following call to the {\em binomial} function in \IDL:
\begin{equation}
P_B = Binomial(C^s; C^s+C^b, \frac{1}{1+A^b/A^s}), \label{PB_eqn}
\end{equation} 
where $C^s$ and $C^b$ are the number of counts observed in the source aperture and background region in a specified energy band.
The source aperture and background region ``areas,'' $A^s$ and $A^b$, are discussed in \S\ref{bkg_extract.sec}.  

This derivation takes into account both the uncertainty of estimating the background level (i.e., the Poisson nature of $C^b$) and the uncertainty of the events observed within the source aperture (i.e., the Poisson nature of $C^s$).
Thus, the significance of a given source extraction will tend to decrease ($P_B$ will rise) if its background region is reduced in size because fewer background counts ($C^b$) will be detected (regardless of whether the background surface brightness inferred from $C^b$ increases or decreases).
This behavior can be intuitively rephrased as ``weak sources benefit more than strong sources from accurate estimates of the background.''
\AEacro\ sizes background regions so that Poisson uncertainty on the background contributes no more than 3\% of the total uncertainty on the source photometry.
When $C^b$ is large and the background is thereby accurately estimated, $P_B$ approaches the familiar integral of the Poisson distribution over the interval [$C^s,\infty$]  \citep[][Appendix~A2]{Weisskopf07}, 
\begin{equation} P_B \simeq 1 - \sum_{i=0}^{C^s - 1} Poisson(i; (A^s/A^b) C^b). \end{equation} 

We recommend using the quantity $P_B$ as the principal measure of the validity of a source's existence, in the context of the iterative source detection strategy described in \S\ref{source_det.sec}. 
$P_B$ can be read as ``the probability that all counts in the source aperture are background'' or ``the probability that no source exists at the extracted location in the presence of the observed local background.''

\section{MODELING THE POINT SPREAD FUNCTION \label{PSFs.sec}}

The \Chandra\ PSF varies both with position on the detector and with energy.  Several models of the \Chandra\ High Resolution Mirror Assembly (HRMA) are available, including \Chart, SAOTrace, \MARX, and the {\it mkpsf} tool in \CIAO.  
\Chart\ has an interactive interface and is thus not suited for automated processing; SAOTrace is available on only a limited set of computer platforms. 
The {\it mkpsf} output has technical limitations such as coarse spatial sampling across the detector, and omission of PSF blurring effects not related to the HRMA.   
We therefore use the \MARX\ ray-trace simulator, running simulations for each observation of each source at five monochromatic energies: 0.277, 1.4967, 4.51, 6.4, and 8.6~keV.\footnote
{
These are the five PSF energies used by the \anchorparen{http://cxc.harvard.edu/ciao/dictionary/psflib.html}{\Chandra\ PSF Library}.
}
\MARX\ dithers the simulated source using the observation's aspect file, allowing accurate modeling of distortions caused by the PSF dithering over the boundaries of the \ACIS\ CCDs.  

On-axis, the PSF of a real source observed by \ACIS\ differs strongly from the HRMA PSF due to three blurring effects.
First, significant quantization effects arise because \ACIS\ pixels do not fully sample the HRMA PSF.
Second, the reported positions of \ACIS\ events are reconstructed from an imperfect aspect solution that measures the dither motion.
Third, the default \CIAO\ pipeline adds a $\pm 0.25$\arcsec\ \anchorfoot{http://cxc.harvard.edu/ciao/why/acispixrand.html}{random number to each event's position.} 
The standard model for these blurring effects is a Gaussian blurring function built into \MARX.\footnote
{
Both \Chart\ and SAOTrace generate ``rays'' from the HRMA, and recommend using \MARX\ to model all three blurring effects.
}
Calibration of this blurring model (the standard deviation of the Gaussian) is available only for all three blurring effects combined; this is not suitable for our needs because we choose to remove the event position randomization during our event pre-processing (\S\ref{L1_L2.sec}). 

Thus, we choose to run \MARX\ simulations with its post-HRMA blurring model disabled, and then blur the simulated event positions ourselves in two steps.
First, \ACIS\ quantization is modeled by convolving the HRMA PSF image with a ``box kernel'' sized to match the \ACIS\ pixel.  We feel a Gaussian kernel, which has infinite extent, is an inappropriate model for pixel quantization (and for the event position randomization, when present).
Second, \anchorfoot{http://asc.harvard.edu/cal/ASPECT/img_recon/report.html}{aspect reconstruction errors} are modeled by convolving the PSF image with a two-dimensional Gaussian kernel with $\sigma_x = \sigma_y = 0.07$\arcsec.




\newpage


\begin{deluxetable}{rcrrrrrrrrrrrccrr}
\centering \rotate \tabletypesize{\tiny} \tablewidth{0pt}

\tablecaption{ Sample \AEacro\ Output Table:  Basic Source Properties \label{src.tab}}

\tablehead{
\multicolumn{2}{c}{Source} &
\multicolumn{4}{c}{Position} &
\multicolumn{5}{c}{Extraction} &
\multicolumn{6}{c}{Characteristics} \\
                                
\multicolumn{2}{c}{\hrulefill} &  
\multicolumn{4}{c}{\hrulefill} &
\multicolumn{5}{c}{\hrulefill} &
\multicolumn{6}{c}{\hrulefill} \\

\colhead{Seq. No.} & \colhead{CXOU J} &
\colhead{$\alpha$ (J2000.0)} & \colhead{$\delta$ (J2000.0)} & \colhead{Error} & \colhead{$\theta$} &
\colhead{$C_{t,net}$} & \colhead{$\sigma_{t,net}$} & \colhead{$B_{t}$} & \colhead{$C_{h,net}$} & \colhead{PSF Frac.} &   
\colhead{Signif.} & \colhead{$\log P_B$} & \colhead{Anom.} & \colhead{Var.} &\colhead{Eff. Exp.} & \colhead{$E_{median}$}  \\

\colhead{} & \colhead{} &
\colhead{(\arcdeg)} & \colhead{(\arcdeg)} & \colhead{(\arcsec)} & \colhead{(\arcmin)} &
\colhead{(counts)} & \colhead{(counts)} & \colhead{(counts)} & \colhead{(counts)} & \colhead{} &
\colhead{} & \colhead{} & \colhead{} & \colhead{} & \colhead{(ks)} & \colhead{(keV)}
 \\

\numberthecolumn & \numberthecolumn & 
\numberthecolumn & \numberthecolumn & \numberthecolumn & \numberthecolumn & 
\numberthecolumn & \numberthecolumn & \numberthecolumn & \numberthecolumn & \numberthecolumn & 
\numberthecolumn & \numberthecolumn & \numberthecolumn & \numberthecolumn & \numberthecolumn & \numberthecolumn 
\setcounter{column_number}{1}
}

\startdata
   1 & 104021.42$-$594810.1 &160.051783 & -59.802827 &     0.0 &     3.0 &  416.8 &    21.0 &     1.2 &    13.7 &    0.89 &   19.4 & $<$-5 & .... & b &    59.8 &     1.0 \\
   2 & 104013.71$-$594643.9 &160.132132 & -59.778874 &     0.1 &     3.2 &  131.1 &    12.0 &     0.9 &     1.6 &    0.90 &   10.5 & $<$-5 & .... & a &    63.3 &     1.0 \\
   3 & 104121.33$-$595825.0 &160.301395 & -59.973638 &     0.4 &     6.3 &   23.8 &     6.0 &     6.2 &     0.0 &    0.89 &    3.6 & $<$-5 & .... & b &    58.8 &     1.0 \\
   4 & 104171.50$-$594037.0 &160.322940 & -59.676954 &     0.0 &     4.9 &  664.9 &    28.1 &    27.1 &   152.3 &    0.40 &   23.0 & $<$-5 & .... & a &    58.3 &     1.4 \\
   5 & 104153.44$-$593945.6 &160.397701 & -59.662688 &     0.2 &     3.7 &   13.1 &     4.3 &     0.9 &     0.0 &    0.90 &    2.7 & $<$-5 & .... & a &    61.3 &     1.1 \\
\enddata

\tablecomments{
\\Col.\ (1): X-ray catalog sequence number, sorted by RA.
\\Col.\ (2): IAU designation.
\\Cols.\ (3) and (4): Right ascension and declination (in decimal degrees) for epoch J2000.0.
\\Col.\ (5): Estimated standard deviation of the random component of the position error, $\sqrt{\sigma_x^2 + \sigma_y^2}$.  The single-axis position errors, $\sigma_x$ and $\sigma_y$, are estimated from the single-axis standard deviations of the PSF inside the extraction region and the number of counts extracted.
\\Col.\ (6): Off-axis angle.
\\Cols.\ (7) and (8): Net counts extracted in the total energy band (0.5--8~keV); average of the upper and lower $1\sigma$ errors on col.\ (7).
\\Col.\ (9): Background counts expected in the source extraction region (total band).
\\Col.\ (10): Net counts extracted in the hard energy band (2--8~keV).
\\Col.\ (11): Fraction of the PSF (at 1.497 keV) enclosed within the extraction region. A reduced PSF fraction (significantly below 90\%) may indicate that the source is in a crowded region. 
\\Col.\ (12): Photometric significance computed as net counts divided by the upper error on net counts. 
\\Col.\ (13): Logarithmic probability that extracted counts (total band) are solely from background.  Some sources have $P_B$ values above the 1\% threshold that defines the catalog because local background estimates can rise during the final extraction iteration after sources are removed from the catalog.
\\Col.\ (14):  Source anomalies: (g) fractional time that source was on a detector (FRACEXPO from {\em mkarf}) is $<$0.9.
\\Col.\ (15): Variability characterization based on K-S statistic (total band): (a) no evidence for variability ($0.05<P_{KS}$); (b) possibly variable ($0.005<P_{KS}<0.05$); (c) definitely variable ($P_{KS}<0.005$).  No value is reported for sources with fewer than four counts or for sources in chip gaps or on field edges.
\\Col.\ (16): Effective exposure time: approximate time the source would have to be observed at the aimpoint of the ACIS-I detector in Cycle ??? to obtain the reported number of net counts (see \url{http://asc.harvard.edu/cgi-bin/build_viewer.cgi?ea}).
\\Col.\ (17): Background-corrected median photon energy (total band).}

\end{deluxetable}

\begin{deluxetable}{rcrrlllrrrrrc}        

\centering \rotate \tabletypesize{\scriptsize} \tablewidth{0pt}

\tablecaption{ Sample \AEacro\ Output Table:  Spectral Fits \label{spec.tab}}

\tablehead{
\multicolumn{4}{c}{Source\tablenotemark{a}} &
\multicolumn{3}{c}{Spectral Fit\tablenotemark{b}} &      
\multicolumn{5}{c}{X-ray Luminosities\tablenotemark{c}} &
\colhead{Notes\tablenotemark{d}} \\ 

\multicolumn{4}{c}{\hrulefill} &
\multicolumn{3}{c}{\hrulefill} &       
\multicolumn{5}{c}{\hrulefill} \\

\colhead{\#} & \colhead{CXOU J} & \colhead{$C_{t,net}$} & \colhead{Signif.} &
\colhead{$\log N_H$} & \colhead{$kT$} & \colhead{$\log EM$} &  
\colhead{$\log L_s$} & \colhead{$\log L_h$} & \colhead{$\log L_{h,c}$} & \colhead{$\log L_t$} & \colhead{$\log L_{t,c}$} &
\colhead{}  \\

\colhead{} & \colhead{} & \colhead{} & \colhead{} &
\colhead{(cm$^{-2}$)} & \colhead{(keV)} & \colhead{(cm$^{-3}$)} & 
\multicolumn{5}{c}{(erg s$^{-1}$)} &
\colhead{} \\

\numberthecolumn & \numberthecolumn & \numberthecolumn & \numberthecolumn & 
\numberthecolumn & \numberthecolumn & \numberthecolumn & 
\numberthecolumn & \numberthecolumn & \numberthecolumn & \numberthecolumn & \numberthecolumn & 
\numberthecolumn 
\setcounter{column_number}{1}
}



\startdata
   1 & 104021.42$-$594810.1 &   416.8 &    19.4 &$21.5$                 & ${\phn}0.84$                 & $ 54.3$                  &  31.47 &   30.32 &   30.34 &   31.50 &   31.82 &2T \\
   2 & 104013.71$-$594643.9 &   131.1 &    10.5 &$20.6_{\cdots}^{+1.3}$ & ${\phn}0.64_{-0.16}^{+0.10}$ & $ 53.8_{-0.2}^{+1.5}$    &  30.97 &   29.53 &   29.53 &   30.99 &   31.03 &\nodata \\
   3 & 104121.33$-$595825.0 &    23.8 &     3.6 &$21.8_{-0.8}^{+0.4}$   & ${\phn}0.47_{-0.3}^{+0.2}$   & $ 53.7_{-0.5}^{+2.0}$    &  30.22 &   28.95 &   28.99 &   30.25 &   30.86 &\nodata \\
   4 & 104171.50$-$594037.0 &   664.9 &    23.0 &$22.1_{-0.11}^{+0.08}$ & ${\phn}0.64_{-0.08}^{+0.07}$ & $ 55.4_{-0.3}^{+0.2}$    &  31.87 &   31.89 &   31.94 &   32.18 &   32.77 &2T \\
   5 & 104153.44$-$593945.6 &    13.1 &     2.7 &$21.9_{-0.5}^{+0.3}$   & ${\phn}0.43_{-0.2}^{+0.2}$   & $ 53.7_{-0.6}^{+0.9}$    &  29.92 &   28.68 &   28.75 &   29.94 &   30.77 &\nodata \\
\enddata

\tablenotetext{a}{ 
For convenience, cols.\ (1)--(4) reproduce the source identification, net counts, and photometric significance data from Table~\ref{src.tab}.
}

\tablenotetext{b}{
All fits used the {\em source} model ``tbabs*vapec'' in XSPEC.  
Cols.\ (5) and (6) present the best-fit values for the extinction column density and plasma temperature parameters.
Col.\ (7) presents the emission measure derived from the model spectrum, assuming a distance of 2.3~kpc. 
Quantities marked with an asterisk (*) were frozen in the fit.  
Uncertainties represent 90\% confidence intervals.
More significant digits are used for uncertainties $<$0.1 in order to avoid large rounding errors; for consistency, the same number of significant digits is used for both lower and upper uncertainties.
Uncertainties are missing when XSPEC was unable to compute them or when their values were so large that the parameter is effectively unconstrained.  
Fits lacking uncertainties, fits with large uncertainties, and fits with frozen parameters should be viewed merely as splines to the data to obtain rough estimates of luminosities; the listed parameter values are not robust.  
}

\tablenotetext{c}{ X-ray luminosities derived from the model spectrum are presented in cols.\ (8)--(12): (s) soft band (0.5--2 keV); (h) hard band (2--8 keV); (t) total band (0.5--8 keV).  
Absorption-corrected luminosities are subscripted with a $c$.
Cols. (8) and (12) are omitted when $\log N_H > 22.5$~cm$^{-2}$ since the soft band emission is essentially unmeasurable.  
}

\tablenotetext{d}{``2T'' means a two-temperature model was used; 
the second temperature is shown in parentheses.   
 }

\end{deluxetable}

\end{document}